\documentclass[letterpaper,11pt]{article}
\pdfoutput=1 

\usepackage{jheppub} 

\usepackage[T1]{fontenc} 
\usepackage{tikz}
\usepackage{dsfont}
\usepackage{amsmath}
\def\ba{\begin{align}}\def\ea{\end{align}}
\def\beq{\begin{eqnarray}}\def\eeq{\end{eqnarray}}
\def\be{\begin{equation}}\def\ee{\end{equation}}
\def\ben{\begin{equation}}
\def\een{\end{equation}}
\def\bea{\begin{eqnarray}}
\def\eea{\end{eqnarray}}
\def\bA{{\bar{A}}}

\def\hM{{\hat{M}}}
\def\cC{\hat{\cal C}}
\def\ccC{{\cal C}}

\def\hpiM{{\hat{\Pi}_M}}
\def\hpi{{\hat{\pi}}}
\def\hPi{{\hat{\Pi}}}
\def\hC{\hat{C}}
\def\hP{\hat{P}}
\def\hX{\hat{X}}
\def\hZ{\hat{Z}}
\def\hR{\hat{R}}

\def\hV{\hat{V}}
\def\hW{\hat{W}}
\def\hQ{\hat{Q}}
\def\hs{\hat{s}}
\def\hl{\hat{\lambda}}
\def\hr{\hat{r}}
\def\hrho{\hat{\rho}}
\def\cO{{\cal{O}}}
\def\cD{{\cal{D}}}
\def\cR{{\cal{R}}}
\def\tX{{\tilde{X}}}
\def\tO{{\tilde{O}}}

\def\trho{{\tilde{\rho}}}

\def\lan{\langle}
\def\ran{\rangle}
\def\mH{{\mathcal H}}
\def\bA{{\bar A}}
\def\eq#1{(\ref{#1})}

\def\eps{\epsilon}
\def\tr{{\rm Tr}}

\title{Gauge Invariant Target Space Entanglement in D-Brane Holography}

\author{Sumit R. Das$^1$,}
\author{Anurag Kaushal$^2$,}
\author{Sinong Liu$^1$,}
\author{Gautam Mandal$^2$,}
\author{Sandip P. Trivedi$^2$}

\affiliation{$^1$Department of Physics and Astronomy, University of Kentucky, Lexington, KY 40506, U.S.A.}
\affiliation{$^2$Department of Theoretical Physics, Tata Institute of Fundamental Research, Mumbai 400005, INDIA}

\emailAdd{das@pa.uky.edu}\emailAdd{anurag.kaushal@theory.tifr.res.in}\emailAdd{sinong.liu@uky.edu}\emailAdd{mandal@theory.tifr.res.in}\emailAdd{sandip@theory.tifr.res.in}

\abstract{It has been suggested in https://arxiv.org/abs/2004.00613  that in Dp-brane holography, entanglement in the target space of the D-brane Yang-Mills theory provides a precise notion of bulk entanglement in the gravity dual. We expand on this discussion by providing a gauge invariant characterization of operator sub-algebras corresponding to such entanglement. This is achieved by finding a projection operator which imposes a constraint characterizing the target space region of interest.  By considering probe branes in the Coloumb branch we provide  motivation for why the operator sub-algebras we consider are  appropriate for describing a class of measurements carried out with  low-energy probes in the corresponding bulk region of interest.
We derive expressions for the corresponding Renyi entropies in terms of path integrals which can be directly used in numerical calculations.}

\begin{document}

\begin{flushright}
TIFR-TH/20-48
\end{flushright}

\maketitle
\flushbottom

\section{Introduction}
\label{sec:intro}


In quantum field theory on a fixed space-time background, entanglement between two regions  of space has a well defined meaning in the presence of a UV cutoff and the corresponding entanglement entropy 
provides valuable information about the nature of the quantum state. In quantum gravity this is a tricky issue since space-time is dynamical. This becomes even more tricky in String Theory where the fundamental degrees of freedom are extended objects.  Nevertheless, in a weakly coupled semi-classical regime there is an approximate notion which comes from thinking of gravity as a field theory of gravitons in a background. It is therefore interesting to ask if there is a precise notion of entanglement in a complete theory of gravity which reduces to the above approximate notion in the appropriate regime.

In \cite{Das:2020jhy} four of us proposed that in gravitational theories which have holographic duals, such as the ones which arise in String Theory, such a precise notion indeed exists. The proposal is that in Dp brane holography for $p < 3$ this notion is provided by entanglement in the target space of the Yang-Mills theory on the brane. The idea is that a suitable target space constraint can be associated with a co-dimension one spatial region in the bulk dual. In the Yang-Mills theory the target space constraint then  leads to a sub-algebra of operators. The expectation values of operators in this sub-algebra  can be obtained correctly using a reduced density matrix lying in the sub-algebra itself. The von-Neumann entropy for this reduced density matrix is  the precise notion, sought above, of a geometric entropy of the bulk sub-region. When the entire system is in a pure state this entropy is entirely  due to quantum entanglement. When the system is in a mixed state, this contains a classical piece. 

We conjectured that when the state is the $N$ brane bound state or its slightly heated up version, this von Neumann entropy is given by the Bekenstein formula $A/4G$ where $A$ is the area of the entangling surface in the dual black brane geometry \footnote{For mixed states a spatial Bekenstein bound is not generally valid. In such situations one may need to consider the fine grained entropy of the degrees of freedom of a light sheet associated with a boundary of area $A$ \cite{bousso}} .  In a complete theory of gravity one would expect that in any definition of geometric entanglement entropy the UV cutoff is automatic. Our conjecture therefore implies that the UV cutoff is provided by Netwon's constant, and not by e.g. the string length. Indeed, for simple entangling surfaces for $p < 3$ our conjecture yields answers which scale as $N^2$ and are expressible purely in terms of the appropriate dimensionless quantities of the Yang-Mills theory - as one would expect. 

The bulk entanglement we are considering is across {\em any} codimension two surface. This is distinct from the corrections to holographic entanglement entropy \cite{Ryu:2006bv, Hubeny:2007xt} due to entanglement across extremal surfaces \cite{Faulkner:2013ana,recentbulk} or what would become a quantum extremal surface \cite{Engelhardt:2014gca}. { Note that the proposal discussed above, for Dp branes with $p > 0$,  considers the same target space constraint  holding at all points 
on   the Dp brane. The boundary of the corresponding codimension one spatial surface in the bulk then includes the entire spatial boundary of  space-time. For $p>0$ it is also possible to discuss a more general notion of entanglement  which arises  for a sub algebra of observables tied to  a target space constraint  that  applies only to a part of the base space along which the Dp branes extend.

The fact that Newton's constant is the natural cutoff is consistent with the idea that the Bekenstein formula for black hole entropy involves a renormalized Newton constant
\cite{,thooft,susskind1,susskind2}. Furthermore it has been argued in \cite{induced} that this naturally happens in theories of induced gravity.
In the past, \cite {jacobson} has argued that Einstein's equations follow from thermodynamics, provided the cutoff in the entanglement entropy in a theory of gravity is Newton's constant. \cite{myers} had also conjectured that the entanglement entropy across an arbitrary surface in a theory of gravity saturates the Bekenstein bound. The reasoning of \cite{Das:2020jhy} is initimately tied with the identification of bulk entanglement with target space entanglement and therefore differs from these other papers in an essential way.

The appearance of Newton's constant as the UV cutoff is also consistent with the calculation of the bulk entanglement entropy in the c=1 Matrix Model / 2d string theory duality \cite{Das:1995vj,Das:1995jw,Hartnoll:2015fca}. The holographic theory is now gauged quantum mechanics of a single $N \times N$ hermitian matrix. In this case the space of eigenvalues can be interpreted as a bulk space and the only propagating mode of the two dimensional string is related to the density of eigenvalues or the collective field \cite{Das:1990kaa} \footnote{The massless mode of two dimensional string theory is related to the density of eigenvalues by an integral transform with a kernel whose scale is the string scale \cite{polnat}. Strictly speaking, the entanglement entropy calculated here is in the eigenvalue space. However this would agree with a bulk notion in terms of usual string theory upto an uncertainty of the order of the string scale.} . The Matrix Model is described exactly by $N$ free non-relativistic fermions in an inverted harmonic oscillator potential which can be rewritten as a second quantized field theory living in the eigenvalue space \cite{Sengupta:1990bt}. Entanglement of a region of the eigenvalue space can be then defined in the usual way in this field theory.  In fact, the c=1 theory provides the simplest example of target space entanglement, since the emergent space is the target space of the matrix model. In an approriate limit, the entanglement entropy agrees with what one would expect from the low energy effective field theory, but with the UV cutoff replaced by the position dependent string coupling. After incorporating the appropriate factor of $N$, the UV cutoff is identified with Newton's constant.

The target space entanglement explored in  \cite{Das:2020jhy} is in a gauge fixed version of the holographic theory. This involves the temporal gauge for the gauge field, and a further gauge choice. For the c=1 model the latter is the gauge where the single matrix is diagonal. The remaining symmetries are Weyl transformations which permute the eigenvalues. In the Dp brane theories we have multiple matrices, and the remaining gauge freedom in the temporal gauge can be used to diagonalize a single matrix which needs to be chosen to express a desired target space constraint. 
The situation studied in detail in \cite{Das:2020jhy} involves diagonalization of one of the scalar fields. The  target space constraint is then expressed in terms of an allowed range of the eigenvalues, e.g. requiring the eigenvalue to be larger than some number. This corresponds to a bulk region characterized by one of the transverse coordinates being larger than some value and a spatial co-dimension one planar entangling surface which bounds this region. The full Hilbert space breaks up into a direct sum of superselection sectors characterized by the number of eigenvalues which satisfy the constraint. In each sector, the smaller Hilbert space is a direct product, which allows one to define a reduced density matrix in the usual fashion. Two possible versions of the proposal for a corresponding reduced density matrix were studied. 

While \cite{Das:2020jhy} specified the general properties of the operators belonging to the sub-algebra of operators associated with a given target space constraint, a procedure to obtain such operators in terms of the operators of the matrix theory was not specified.
Furthermore, in this gauge fixed formalism, it is difficult (though not impossible) to describe general entangling surfaces.

In this paper we address both these issues by developing a {\em gauge invariant} description of target space entanglement. This will be achieved by constructing a projection operator appropriate for the desired target space constraint. Starting with a gauge invariant operator which contains a string of matrices, the subalgebra then consists of operators obtained by projecting each of these matrices. We show that in the gauge used in \cite{Das:2020jhy} these yield the correct class of operators in each superselection sector. Moreover, the gauge invariant construction enables us to easily formulate other target space constraints, e.g. those which correspond to entangling surfaces in the bulk at a given value of the radial coordinate in the transverse space of the D-branes - in  
 this case we also show how the target space constraint can be implemented  explicitly by developing a formalism for a polar decomposition of the matrices.

The proposed connection of a target space constraint with a bulk region is based on several ingredients of gauge-gravity duality and closely tied  to  the emergence of bulk locality. As is well known, the velocity dependent potential between two stacks of D0 branes in supergravity follows from an effective action calculation in the Coulomb branch of D0 brane quantum mechanics \cite{dkps,bfss,kt, tram,rseven,twoloop,polbecker}. For example, one may consider a single D0 brane stripped off from a stack of $N$ D0 branes in their bound state, corresponding to a point on the Coulomb branch of the D0 brane matrix theory. The Higgs vev at this point is then the transverse location of this probe D0 brane. This implies that a restriction to a region $\cR$ of the bulk  can be described as a restriction in the target space of the brane theory. In the 't Hooft limit, the gravity dual of the bound state of D0 branes is a non-trivial supergravity background \cite{imsy}, and the velocity dependent potential can be obtained from  a DBI-CS action for  the probe  D0 brane moving in this background. It can also be calculated from the effective action evaluated at the corresponding point in moduli space in the gauge theory. 
Supersymmetry guarantees in fact that the leading terms in the effective action may be calculated perturbatively. We argue that the potential will also agree with the effective action for operators in the subalgebra, ${\cal A}_{\cal R}$, which we associate with the region ${\cal R}$, thereby arguing that the subalgebra contains operators needed to describe bulk measurements which can be carried out in ${\cal R}$.

A somewhat stronger connection comes from a point on the Coulomb branch $SU(N) \rightarrow SU(N-2) \times U(1) \times U(1)$, which corresponds to two D branes stripped off from the rest. This situation has been studied in detail for D3 branes in \cite{branewaves}. In this case, when the two individual branes are excited, the lowest order terms in the effective action of the two $U(1)$'s agree precisely with a supergravity calculation of the potential between the two branes which follow from exchanges of supergravity modes propagating on the $AdS_5 \times S^5$ produced by the remaining $(N-1)$ branes. This agreement is more detailed than the single D0 DBI+CS action since the supergravity modes in this background mix non-trivially. There should be a similar agreement for D0 branes.

Going beyond the ground state, in an excited state of the gauge theory which corresponds to a modified supergravity background, also one expects that the potential experiences by a probe brane can be obtained from the DBI+CS action, and this potential should agree with an effective action calculation which can be carried out in the gauge theory keeping operators in the subalgebra  ${\cal A}_{\cal R}$. Of course a perturbative calculation will no longer suffice to demonstrate this{\footnote{ unless the excited state preserves a high degree of supersymmetry\cite{Douglas:1997pj}}. But one might hope to be able to check this as numerical techniques improve further.
In fact some progress has already been made along these lines in obtaining the dynamics of probe branes  at finite temperatures \cite{hanadaprobe}. In these calculations some evidence was found that the supergravity fields couple  to operators in the probe brane in a manner consistent with the generalized AdS/CFT correspondence discussed in \cite{jevyoneya,sekinoyoneya}.

To summarize, the dynamics on the Coulomb branch   should allow one to measure the local background, at least at the level of one point functions, for  gravity and other supergravity modes, in a region ${\cal R}$. This dynamics we argue can be obtained in the gauge theory   by studying the  effective action for gauge invariant operators. If we  are interested in measuring the supergravity fields {\em only in the region $\cR$ of the bulk}, we argue that it is sufficient to  only consider the operators in the subalgebra ${\cal A}_{\cal R}$ associated with ${\cal R}$. As mentioned above, this subalgebra contains  gauge invariant operators  obtained after carrying out a suitable projection determined by the target space constraint which corresponds to the bulk region ${\cal R}$. 

While the discussion above pertains to the Coulomb branch, the considerations should be valid for a  general {\em configuration} which appears in the wavefunction of the $N$ D0 brane bound state. This motivates our identification of bulk entanglement with target space entanglement.

We should mention that we expect the effective action and the related correlation functions of the projected operators to provide only some and not {\it all} of  the detailed information about  supergravity modes and the dual boundary operators related to them via the BDHM-HKLL construction \cite{hkll}. In particular, the energy momentum tensor is not contained in the sub-algebra, only its projected version is. We expect  that  this  imposes important  limitations on the extent to which we can learn about the stress energy tensor's correlation functions from the sub-algebra. In fact, as was importantly argued in  \cite{suvrat,suvrat2}, if  the sub-algebra would allow all information pertaining to the stress tensor to be obtained, then for an annular region adjacent to the boundary, the entanglement entropy would be exactly zero \footnote{The argument is as follows:  if the energy-momentum tensor at all points on the boundary is included in the set of observables, so is the energy and the projector to the ground state. In the vacuum,  the latter is the density matrix of the whole system. This would mean that the associated entanglement entropy must be exactly zero. In our case the energy, and therefore also the  ground state projector, is not an element  of the sub-algebra.}. In this sense the association of a target space constraint with a bulk region is approximate.

An analytic calculation of the target space entanglement entropy requires an explicit expression for the wavefunction. Even in the simplest case of D0 branes, explicit expressions for the bound state wavefunction is not known, though the the existence of bound states of D0 branes has been proven \cite{bound} \footnote{For bosonic BFSS models, the existence of bound states has been proved both numerically \cite{Kawahara:2007fn} and analytically in the limit of large dimensions \cite{Mandal:2009vz}.}. There are candidates for approximate wavefunctions which can be in principle used to perform analytic computations of the von Neumann entropy \cite{approx}. However, there has been substantial progress in numerical calculations of properties of D0 brane bound states at finite temperature: these calculations provide precision tests of the AdS/CFT correspondence \cite{catterall,Hanada:2016zxj}. These calculations deal with thermodynamic quantities, correlation functions \cite{hanadayoneya} and investigations of probe dynamics \cite{hanadaprobe}. In this paper we derive path integral expressions for target space Renyi entropies which can be directly used to perform numerical calculations. Work in this direction is being developed currently \cite{ushanada}: these calculations should prove or disprove our conjecture about saturation of the Bekenstein bound.

The formalism of target space entanglement entropy has been developed in \cite{DMT:2018} and \cite{Mazenc:2019ety}. Notions similar to target space entanglement have been used to define entanglement in string theory in the worldsheet formalism \cite{stringee} and in various explorations in holographic entanglement \cite{internalextremal}. 
Another notion of entanglement of internal degrees of freedom (also combined with spatial degrees of freedom) called entwinement has been discussed in \cite{entwinement}.
Notions of entanglement associated with other kinds of partitions of large-N degrees of freedom have been explored in \cite{alet}. The proposal of \cite{Das:2020jhy} is distinct from these other works.

The paper \cite{vanrams} has explored general {\em extremal} surfaces in D brane geometries (as distinct from RT surfaces) and speculated on possible meanings of their areas with entanglement of degrees of freedom in the D0 brane quantum mechanics.  In particular, these authors have considered subsets of operators consisting of linear combinations of traceless symmetric products of the matrices in the D0 brane theory which would correspond to functions which have support on some region of $S^8$ and speculated that an entropy can be associated with such a subset. Our proposal is quite different from this: we aim to describe bulk entanglement which involves the radial direction as well, and we associate an entropy with a closed subalgebra.

This paper is organized as follows. 
 In section \ref{gaugeinv} we introduce the gauge invariant construction of operator algebras which define a target space entanglement. We show how this construction leads to the gauge fixed version discussed in \cite{Das:2020jhy} and review the proposed connection to bulk entanglement and our conjecture about the saturation of Bekenstein bound. We also discuss how to impose radial constraints in target space by developing a polar decomposition of matrices. In section \ref{tsebe} we discuss the connection of target space entanglement and bulk entanglement. In section \ref{tsebb} we recapitulate the conjecture in \cite{Das:2020jhy} that the target space entanglement entropy saturates the Bekenstein bound. In section \ref{piere} we derive path integral expressions for target space Renyi entropies which can be directly used for numerical calculations. Section \ref{cncl} contains concluding remarks. The  Appendix \ref{appone} contains some details of the construction of projected operators. Appendix \ref{apptwo} provides details of matrix polar decompositions for multiple matrices. Appendix \ref{appthree} deals with the DBI+CS action of a single D0 brane in the supergravity background produced by $N$ other extremal branes and its comparison with D0 brane quantum mechanics effective action.

\section{Gauge Invariant Target Space Entanglement}
\label{gaugeinv}

In this section we will show how target space entanglement in a theory of multiple matrices can be formulated in a gauge invariant fashion. A more detailed description appears in Appendix \ref{appone}.

\subsection{Review of the gauge-fixed formulation}\label{gaugeinv2.1}

In a previous paper \cite{Das:2020jhy}, we considered the $D0$ brane theory and discussed a bulk region specified by a condition on one of the  spatial bulk coordinates,  say $x^1$.  The  condition took the form,
\ben
x^1>a,
\label{bulk-region}
\een for some real number $a$. We proposed that this condition mapped to a target space constraint in the  quantum mechanical dual  theory that lives on the boundary. And the bulk entanglement entropy maps to the entanglement entropy associated with this target space constraint in the boundary theory. The  entanglement entropy defined in this way is manifestly finite when  $N$ is finite. 

The action of D0 brane quantum mechanics is given by
\ben
S = \frac{N}{2 (g_s N) l_s}{\rm Tr}\int dt  \left[ \sum_{I=1}^9 (D_tX^I)^2  - \frac{1}{ l_s^4} \sum_{I\neq J = 1}^9 [ X^I, X^J]^2 \right] + {\rm fermions}
\label{three-one}
\een
where $X^I$ are $N \times N$ hermitian matrices, and the covariant derivative is defined by
\ben
D_t X^I \equiv \partial_t X^I + i[A_t,X^I]
\label{two}
\een
In the example above, the target space constraint  involves the operator $X^1$ in the boundary theory.  To specify the target space constraint, we worked in the gauge where $A_t =0$. The remaining gauge freedom consists of time independent $SU(N)$ rotations, which we fixed by requiring $X^1$ to be diagonal. The corresponding operators and their canonical conjuagte momenta have the form
\bea
\hX^1 & \rightarrow  & {\rm diag} \left( \hl_1,\cdots \hl_N \right) \nonumber \\
\hPi_1 & \rightarrow & {\rm diag}\left( \hpi_1  \cdots \hpi_N  \right)
\label{four-four}
\eea
This does not fix the gauge completely: we are left with Weyl transformations,
\bea
(\hl_1,\hl_2, \cdots ,\hl_N) & \mapsto & (\hl_{\sigma(1)},\hl_{\sigma(2)}, \cdots ,\hl_{\sigma(N)}),\, \sigma \in S(N) \nonumber \\
\hX^L & \mapsto &
\sigma(\hX^L),\; \sigma(\hX^L_{ij})= \hX^L_{\sigma(i)\sigma(j)}, \, L=2, \cdots ,9.
\label{permute-x}
\eea
and $U(1)^N$ transformations which keep the diagonal matrix elements of all the matrices invariant and multiplies the off-diagonal elements by phases \footnote{In a previous version of this paper which appeared on the arXiv, we  did not consider these $U(1)^N$ transformations. Subsequently the paper \cite{lawrence} appeared, where these remaining symmetries were emphasized.}
\ben
\hX^L_{ij} \mapsto \hX^L_{ij} e^{i(\theta_i - \theta_j)}
\label{u1}
\een
where $\theta_i$ are angles. The physical state is constructed by adding Weyl and $U(1)^N$ transforms. Let us work in a basis in the Hilbert space comprised of eigenvectors of the operators $\hl_i,\hX^L_{ij}$ with eigenvalues $\lambda_i, X^L_{ij}$. As is well known the transformation to the eigenvalues of $X^1$ leads to a van der Monde factor in the measure of integration. In the following we will absorb a square root of this factor in the wavefunction so that the modified wavefunction is antisymmetric under an interchange of the eigenvalues. Then a Weyl and $U(1)^N$ symmetrized state is \footnote{Our conventions for normalization of states is different from \cite{Das:2020jhy}.}$^,$\footnote{Since the Weyl group elements $g_W$ and the $U(1)^N$ group elements $g_U$ do not commute, the combined action on a given state $|\psi \rangle$ depends on the order in which the group elements act. In \eq{four-eleven} we have applied $g_U$ followed by $g_W$. It is not difficult to see, however, the `symmetrized' state, which involves sum over the entire set of transforms, does not depend on the order: $\sum_{W,U} g_W g_U \psi = \sum_{W,U} g_U g_W\psi $.}
\ben
| \{ \lambda_i \}, \{ X^L_{ij} \} \rangle_W = \frac{1}{N !}\int \prod_{I=1}^N \frac{d\theta_i}{2\pi} \sum_{\sigma \in S_N}{\rm sgn} (\sigma)
| \{ \lambda_{\sigma(i)} \}, \{ X^L_{\sigma(i) \sigma(j)} e^{i(\theta_{\sigma(i)} - \theta_{\sigma(j)})} \} \rangle
\label{four-eleven}
\een
Note that this symmetrized state is not an eigenstate of the individual $\hl_i,\hX^L_{ij}$'s. They are eigenstates of gauge invariant operators which are traces of products of $\hX^I,\hPi_I$ or products of these traces.

In this gauge, it was proposed that the required target space condition,
corresponding to \eq{bulk-region}, on an eigenvalue $\lambda_i$, was
\be
\label{conda}
\lambda_i>a. 
\ee
The target space constraint can be generalized trivially to 
\ben
\lambda_i \in A
\een
where $A$ is some interval on the real line, with a corresponding change in the bulk region \eq{bulk-region}.

Since there are $N$ eigenvalues the constraint gives rise to $N+1$ different possibilities depending on whether  $0, 1, \cdots,  N,$  of the eigenvalues meet the constraint. These different possibilities actually can be thought of as giving rise to different superselection sectors. The Hilbert space thus becomes a direct sum of Hilbert spaces,
\ben  
\mH_N = \oplus_{k} \mH_{k,N-k}
\label{four}
\een 
where $\mH_{k,N-k}$ denotes the sector where $k$ of the eigenvalues of $\hX^1$ are in the region of interest $A$ and the rest in its complement $\bA$.

The reduced density matrix in the $k^{\rm th} $ superselection sector, $\trho_{k,N-k}$, can be obtained by tracing out the degrees of freedom corresponding to the remaining $(N-k)$ eigenvalues. The corresponding target space entanglement entropy can then  be obtained as the von Neumann entropy for this density matrix and  the full entanglement entropy for all sectors can be obtained by adding the entropy from each sector \footnote{The sector with no eignvalues meeting the required condition is important to keep in mind. It is taken to be one dimensional and the density matrix is then a number corresponding to the probability of finding no eigenvalue meeting the constraint.}. Note that the density matrix in each individual sector is not normalized. Rather the trace ${\rm tr}_{k} \trho_{k,N-k}$ is simply the probability that $k$ of the eigenvalues are in the region of interest. The full reduced density matrix is block diagonal, where each block corresponds to a superselection sector
\ben
\label{fullrho}
\rho = \begin{pmatrix}
\trho_{0,N} & {\bf 0} & {\bf 0} &\cdots & {\bf 0} \\
{\bf 0} & \trho_{1,N-1} & {\bf 0} & \cdots & {\bf 0} \\
\cdots & \cdots & \cdots & \cdots & \cdots \\
{\bf 0} & {\bf 0} & {\bf 0} & {\bf 0} & \trho_{N,0}
\end{pmatrix}
\een
does have unit trace, so that 
\ben
S = - {\rm tr}( \rho \log \rho  ) = -\sum_{k=0}^N {\rm tr}_{k}( \trho_{k,N-k} \log \trho_{k,N-k}  )
\label{totalentropy}
\een
is a legitimate von Neumann entropy.

Actually our proposal had two versions which arise when we think more precisely about tracing  out the degrees of freedom corresponding 
to the remaining $N-k$ eigenvalues. By a suitable choice of gauge the eigenvalues of $X^1$ meeting the constraint can be taken to be the first $k$ eigenvalues. 

In the rest of the paper, the matrix indices $i,j,\cdots = 1 \cdots N$; the indices $a,b = 1 \cdots k$ and $\alpha, \beta = k+1 \cdots N$. In the rest of this subsection, we denote $X^2, X^3 , ..., X^9$ by $X^L$, $L=2,3,...,9$.

\begin{enumerate}

\item{} In the first version, one  traces  out the degrees of freedom corresponding to the $(N-k)$ eigenvalues of $X^1$ which do not satisfy eq.(\ref{conda}), $\lambda_\alpha$ and the degrees of freedom in $(N-k) \times (N-k) $ block  for the remaining spatial matrices, $X_{\alpha\beta}^2, X_{\alpha\beta}^3, \cdots X_{\alpha\beta}^9$. In addition, one also traces out the degrees of freedom corresponding to the off-diagonal elements $(X^2)_{a\alpha}, (X^2)_{\alpha a}$; and similarly for $X^3,X^4, \cdots X^9$. As a result the only degrees of freedom we retain are in the $k \times k$ block. In the basis we are using, the reduced density matrix for the $k$'th sector is then given by
\begin{align}
\trho^{(1)}_{k,N-k}  & \left(  \lambda_a, X^L_{ab}; \lambda^\prime_a, X^{\prime L}_{ab} \right) =  \nonumber \\
& {N \choose k} \int [d\lambda_\alpha dX^L_{a\alpha}dX^L_{\alpha a}dX^L_{\alpha\beta}]~\rho_{tot} \left(\lambda_a,X^L_{ab},\lambda_\alpha, X^L_{a\alpha},X^L_{\alpha a} ,X^L_{\alpha \beta}; \lambda^\prime_a,, X^{\prime L}_{ab}, \lambda_\alpha, X^L_{a\alpha}X^L_{\alpha a} X^L_{\alpha \beta} \right)
\label{three-nine}
\end{align}
where $\rho_{tot}$ is the density matrix of the state of the entire system.

\item{} In the second version,  one  traces out only the degrees of freedom which lie in the $(N-k)\times (N-k)$ blocks for all matrices and retains the remaining degrees of freedom. So for $X^1$ which is diagonal we retain the first $k$ eigenvalues which meet the constraint, but for $X^2$  we retain not only the the elements $(X^2)_{ab}$ but also the off-diagonal elements $(X^2)_{a\alpha}, (X^2)_{\alpha a}$ and similarly for $X^3, \cdots X^9$,
\begin{align}
\trho^{(2)}_{k,N-k}  & \left(  \lambda_a, X^L_{ab}, dX^L_{a\alpha} dX^L_{\alpha a}; \lambda^\prime_a, X^{\prime L}_{ab} dX^{\prime L}_{a\alpha} dX^{\prime L}_{\alpha a} \right) =  \nonumber \\
& {N \choose k}\int [d\lambda_\alpha dX^L_{\alpha\beta}]~
\rho_{tot} \left(\lambda_a,X^L_{ab},\lambda_\alpha, X^L_{a\alpha},X^L_{\alpha a} ,X^L_{\alpha \beta}; \lambda^\prime_a,, X^{\prime L}_{ab}, \lambda_\alpha, X^{\prime L}_{a\alpha}X^{\prime L}_{\alpha a} X^L_{\alpha \beta} \right)
\label{three-nine-1}
\end{align}

\end{enumerate}

In each sector labelled by $k$, the corresponding density matrix evaluates expectation values of a closed subalgebra of operators which correspond to measurements on the variables which are retained. In the first version, the action of such an operator on a general Weyl and $U(1)^N$ symmetrized state of the form (\ref{four-eleven}) has the form
\begin{align}
{\hat{\cO}}~  & |\{ \lambda_a, X^L_{ab},X^L_{a\alpha},X^L_{\alpha\beta},\lambda_\alpha \} \rangle_W \nonumber \\
& = \int [d\lambda_a^\prime ] [dX^{\prime L}_{ab}] ~ \tO ( \{ \lambda_a^\prime,X^{\prime L}_{ab})\},\{\lambda_a,X^L_{ab}\} ) ~|\{ \lambda_a^\prime X^{\prime L}_{ab} \}; \{ \lambda_\alpha,X^L_{a\alpha},X^L_{\alpha\beta} \} \rangle_W
\label{fiveb}
\end{align}
Operators which satisfy this form a subalgebra:
$\tO ( \{\lambda_a,X^L_{ab}\} , \{ \lambda_a^\prime,X^{\prime L}_{ab})\} )$ then denote the matrix elements of an operator in the smaller Hilbert space in this sector.
The reduced density matrix which evaluates expectation values of such operators is given by (\ref{three-nine}).

Similarly for the second version the action is given by
\begin{align}
{\hat{\cO}}~ & |\{ \lambda_a, X^L_{ab},X^L_{a\alpha},X^L_{\alpha\beta},\lambda_\alpha \} \rangle_W \nonumber \\
&  =  \int [d\lambda_a^\prime ] [dX^{\prime L}_{ab}] [dX^{\prime L}_{a\alpha}] [dX^{\prime L}_{\alpha a}]~  \tO (  \{ \lambda_a^\prime,X^{\prime L}_{ab}, X^{\prime L}_{a\alpha},X^{\prime L}_{\alpha a}
)\}, \{\lambda_a,X^L_{ab},X^L_{a\alpha},X^L_{\alpha a}\}  ) \nonumber \\
& \kern200pt  ~|\{ \lambda_a^\prime X^{\prime L}_{ab} X^{\prime L}_{a\alpha} X^{\prime L}_{\alpha a} \}, \{ \lambda_\alpha X^L_{\alpha\beta} \} \rangle_W
\label{fiveg}
\end{align}
 It is clear that the density matrix is again of the form (\ref{fullrho}). 
 
 The associated entanglement entropy for a density matrix of the form eq.(\ref{fullrho})  is expressible in terms of the {\em normalized} density matrices of the subsectors, ${\hat{\rho}}_{k,N-k} = \frac{1}{p_{k,N-k}} \trho_{k,N-k}$ as 
\ben
S = -\sum_{k=0}^N \left[ p_{k,N-k} \log p_{k,N-k} + p_{k,N-k} {\rm tr}_k ({\hat{\rho}}_{k,N-k} \log {\hat{\rho}}_{k,N-k} ) \right]
\label{entropynormalized}
\een
The distillable part of the entanglement is only the second term in (\ref{entropynormalized}), while the first term is a classical piece which cannot be used as a quantum resource for teleportation \cite{ST,V}.

Before closing this subsection let us note that while we have focussed on bosonic operators above,  a similar discussion also applies to fermionic operators in the theory. Depending on which version  of our proposal we consider, the appropriate adjoint color degrees of freedom for  fermonic operators are also to be retained in the sub-algebra. 

\subsection{Gauge-invariant formulation}\label{gaugeinv2.2}

A drawback of the discussion in the previous paper \cite{Das:2020jhy}, and our discussion above, is that this description of the target space constraint  and the  related entropy has been given in a particular gauge, e.g., for the example above we worked in the gauge where $X^1$ is diagonal. Furthermore, while (\ref{fiveb}) and (\ref{fiveg}) describe the properties satisfied by operators belonging to the relevant subalgebra of observables, this does not tell us {\em what} these operators are in terms of the basic operators of the theory. In this subsection, we will address both these issues and give a gauge invariant description of the target space constraint; this will also allow us to generalise the discussion considerably to a much wider class of bulk regions. 

In general, suppose we have a region in the bulk  at time $t$ specified by one condition among the $9$ spatial coordinates,
\be
\label{condo}
f(x_i)>0
\ee
We would like to specify the target space constraint corresponding to this bulk region  in a gauge invariant manner. For this purpose, instead of starting with  a wave function, constructing the density matrix by a partial trace over some degrees of freedom and calculating its entropy,  it is useful to think of the entanglement entropy as arising because one is dealing with a suitable {\it sub-algrebra} of the set of all observables.  The sub-algebra corresponds to the operators whose expectation values can be obtained correctly  from the reduced density matrix obtained after  tracing out the unwanted degrees of freedom. Specifying the sub algebra is an equivalent way of specifying the tracing out procedure and implementing the target space constraint. 

Note that when we think in this way, starting from a sub-algebra of all observables,  the density matrix  itself must lie  in the sub-algebra of observables and, as mentioned, must give the correct expectation values for all operators in the sub-algebra. In addition the  density matrix  is normalised, as usual, to meet the condition, ${\rm tr}\rho=1$. This specifies the density matrix uniquely and the  entanglement entropy is then the  von-Neumann entropy of this density matrix.\footnote{The uniqueness can be easily seen. Suppose there are two possible density matrices, $\rho$ and $\tilde \rho$, which satisfy ${\rm Tr}(\rho O)$ =${\rm  Tr}(\tilde \rho O)$ = ${\rm Tr}(\rho_{tot} O)$ for {\it all} operators $O$ in the sub-algebra. Hence ${\rm Tr}[(\rho - \tilde \rho) O]=0$ for all such operators; since both $\rho$ and $\tilde \rho$ belong to the sub-algebra, this can only happen if $\rho= \tilde \rho$.}   When there are superselection sectors, as in our current discussion, we found above a corresponding density matrix in each sector. However, as we will see, and this is one of the virtues of  specifying a  sub-algebra  to implement the target space constraint,  the sub-algebra of interest can in fact be {\em specified once and for all} in a gauge invariant manner regardless of the sector we are working in. 

Before proceeding to a gauge invariant formulation let us first address the question; how do we determine the relevant subalgebra of operators even in a fixed gauge. To illustrate the procedure it is useful to consider the simple case of gauged quantum mechanics of a single matrix $\hM$. In the gauge where the matrix is diagonal, this reduces to a theory of $N$ fermions on a line. The position and momentum operators of individual fermions are the $\hl_i$ and $\hpi_i$. Consider a typical one body operator in this theory
\ben
\cC_{n,m} = \sum_{i=1}^N \hl_i^n \hpi_i^m
\label{20-1}
\een
We want to impose a target space constraint where the eigenvalues of $\hl_i$ lie in a certain interval on the line denotes by $A$. This corresponding subalegbra consists of operators which act only on the fermions which lie in this interval. Such an operator can be constructed as follows. Define a projection operator
\ben
(\hP_A)_i = \int_A dx~\delta (x - \hl_i)
\label{20-2}
\een
where $A \subset R$. By considering matrix elements between arbitrary states it is clear that this operator indeed satisfies
\ben
(\hP_A)_i^2 = (\hP_A)_i 
\label{20-3}
\een
Now, starting from an operator (\ref{20-1}) construct an operator by replacing each of the $\hl_i, \hpi_i$ by $(\hP_A)_i \hl_i (\hP_A)_i$ and $(\hP_A)_i \hpi_i (\hP_A)_i$ to get
\ben
(\cC_{n,m})^P_A  =  \sum_{i=1}^N (\hP_A)_i \hl_i^n (\hP_A)_i \hpi_i (\hP_A)_i \hpi_i (\hP_A)_i \cdots (\hP_A)_i \hpi_i (\hP_A)_i 
\label{20-4}
\een
where we have used (\ref{20-3}) and the fact $[(\hP_A)_i, \hl_j ]= 0$ to simplify the expression. It may be now easily checked
that the expectation value of $(\cC_{n,m})^P_A$ in a general many particle state becomes a sum of terms: each term corresponds to a sector with $k$ particles in the interval $A$. The k-th term contains the contribution {\em only from the k particles in A}. This is discussed in more detail in Appendix \ref{appone}.

It is now straightforward to construct these operators in a gauge invariant fashion. In terms of the matrix valued operator $\hM$ the projector is clearly given by
\ben
(\hP_A)= \int_A dx~\delta (x {\bf I} - \hM)
\label{20-5}
\een
where ${\bf I}$ is the $N \times N$ identity operator.This procedure generalizes to the D0 brane theory with multiple matrices as we now describe. 

To obtain a sub-algebra  which corresponds to the  target space constraint following from eq.(\ref{condo})  we consider its target space analogue, 
 \be
 \label{cond2}
 f({\hat X}^I)>0
 \ee
 and  the following projection operator which follows from this constraint
\be
\label{proms}
\hP_1=\int_{x>0}  dx \delta (x{\bf I} - f(\hX^I)) 
\ee
where $\hX^I$ are the operators in D0 brane quantum mechanics.
Note that we have taken the function $f$ here to be  the same as in eq. (\ref{condo}) but its argument in eq.(\ref{proms}) are now operators \footnote{More generally the target space constraint and bulk constraint could be related in a more complicated fashion, see below for further discussion of this point.}.  The integral is over positive values of $x$ which is a $c$ number. We will choose the operator $f(X^I)$ to be hermitian.

In general there will be ordering ambiguities which will arise in going from the function $f$ in the bulk to the corresponding function $f$  of matrix operators which appears  in eq.(\ref{proms}); we will comment on this issue further  towards the end of this subsection.

By doing a unitary transformation and going to a basis in which $f(X^I)$ is diagonal one can easily check that $\hP_1$  is  a projection operator satisfying the condition 
\be
\label{proc}
\hP_1^2=\hP_1
\ee
Gauge invariant operators can now be obtained by conjugating with  $\hP_1$ and taking a trace. For example, starting  from $\hX^I, I=1,\cdots 9,$ we construct the corresponding projected operators  $\hP_1 \hX^I \hP_1, I=1, \cdots 9$, and then take a  trace over the color degrees to obtain gauge invariant operators from these projected operators
$Tr(\hP_1 \hX^1), Tr(\hP_1 \hX^2), \cdots Tr(\hP_1 X^9) $ (here we have used cyclicity of the trace and the facts that $[ \hP_1, \hX^I] = 0$ and $\hP_1^2=1$ to drop one of the two $\hP_1$ factors) . 

More generally let ${\cal  O}$ be any operator obtained by multiplying a string of $\hX^I$'s and $\hPi_I$'s where the $\hPi_I$'s  are the momenta
conjugate to the $\hX^I$'s. Schematically we can write  ${\cal O} = \cdots \hX^I\cdots \hPi_J \cdots $ to depict the string of $X^I$'s and $\hPi_J$'s in some order. 
We can obtain a  gauge invariant operator from ${\cal O}$  by taking the colour trace, 
\be
\label{ctrace}
{\hat {\cO}}=Tr({\cal O}) = Tr(\cdots \hX^I \cdots \hPi_J \cdots).
\ee
Now to obtain elements of the desired sub-algebra we consider the projected operators, 
\be
\label{changever1}
\hX^I \rightarrow (\hX^I)^{P_1} = \hP_1 X^I\hP_1
\ee
 and 
 \be
 \label{changever1b}
 \hPi_J \rightarrow (\hPi_J)^{P_1}= \hP_1 \hPi_J \hP_1
 \ee 
and construct the string 
\be
\label{stringa}
 \cdots  (\hX^I)^{P_1} \cdots ( \hPi_J)^{P_1}  \cdots 
 \ee
by replacing every factor of $\hX^I,\hPi_J$ in ${\cal O}$ above with the projected counterpart. Then the projected  operator corresponding to ${\hat {\cO}}$ is given by 
taking the colour trace of eq.(\ref{stringa}). We  will use the notation ${\hat {\cO}}^{P_1}$ for this operator below, so we have, 
\be
\label{opproj}
{\hat {\cO}}^{P_1}= Tr( \cdots (\hX^I)^{P_1} \cdots ( \hPi_J)^{P_1} \cdots). 
\ee
It is important to note that the operator in eq.(\ref{opproj}) is different from $Tr(P_1 {\cal O}P_1)$,
where ${\cal  O}$ is given by eq.(\ref{ctrace}). E.g., when ${\cal O}$ above is $X^I X^J$, ${\hat {\cO}}= Tr(\hX^I \hX^J)$ and $Tr(P_1 {\cal O}P_1 )= Tr (P_1 \hX^I \hX^J P_1) $.
However the operator $ {\hat {\cO}}^{P_1}= Tr(P_1 \hX^I P_1 \hX^J)$ which is different. 

The full sub -algebra we consider  associated with the constraint eq.(\ref{condo}) involves all  single trace operators obtained after projection in this manner 
 and the  multi trace operators obtained from products of such  single trace  projected operators. 

Actually the projection operator $P_1$ above implements version 1) of the proposal,  for a   constraint specified by the function $f(X^I)$. To see this consider  the case $f(X^I)= X^1-a$, discussed above. Working in the gauge where 
$X^1$ is diagonal, let us consider the sector where the first $k$ eigenvalues $x^1_i>a, i=1, \cdots k$ are in the region of interest. Then it is easy to see in this sector that the operator $P_1$ is the matrix 
\begin{equation}
\label{matrixpi}
P_1=
\begin{pmatrix}
  {\bf I}_{k\times k} &  {\bf 0}_{k \times (N-k)} \\
  {\bf 0}_{(N-k)\times k} & {\bf 0}_{(N-k)\times (N-k)} 
 \end{pmatrix}
\end{equation}
where ${\bf I}_{k \times k}$ denotes the identity in the $k \times k$ block and ${\bf 0}$ denotes a matrix where all entries vanish. 
Projecting with this operator we retain for all matrix operators  their upper left hand   $k\times k$ block, as shown in detail in the Appendix \ref{appone}.  Gauge invariant operators made out of such matrix operators are exactly the observables whose expectation values can be calculated using the density matrix obtained from the tracing out procedure described above for version 1) of the proposal.

More generally, for a constraint  $f(x^I)>0$ we can go to the gauge where $f(\hX^I)$ is diagonal
and in the sector where the first $k$ eigenvalues satisfy the constraint find that 
 multiplying with $P_1$ will retain similarly the upper left hand $k \times k$ block for all operators and thus give the correct sub-algebra associated with version 1). 

Implementing the version 2) proposal in a gauge invariant manner is also similarly doable. We first consider the orthogonal projector 
\be
\label{orthodox}
{\tilde P}_1= \int_{x<0}  dx \delta (x{\bf I} - f(\hX^I)),
\ee
which  involves the same argument for the delta function but with the range of the $x$ integral now lying in the complementary region $x<0$. 
It follows that ${\tilde P}_1^2={\tilde P}_1$ 
To implement version 2) we consider  the operators, $\hX^I$ and retain the elements corresponding to $\hX^I - {\tilde P}_1 \hX^I {\tilde P}_1$,
so that 
\be
\label{chngxi}
\hX^I \rightarrow  (\hX^I)^{P_2}= \hX^I - {\tilde P}_1 \hX^I {\tilde P}_1
\ee
 and  similarly for the momentum operators $\hPi_I, I=1,\cdots 9$,  
 \be
 \label{chngpi}
 \hPi_I \rightarrow (\hPi_I)^{P_2} = \hPi_I-  {\tilde P}_1 \hPi_I {\tilde P}_1   
 \ee   
Then taking the trace of a string of such operators we obtain gauge invariant operators
\be
\label{ver2op}
{\hat \cO}^{P_2}= Tr( \cdots (\hX^I)^{P_2}   \cdots (\Pi_J)^{P_2} \cdots)
\ee
which should be compared with eq.(\ref{opproj}) obtained above for the version 1) case. 
It should be emphasised that the transformation $\hX^I \rightarrow (\hX^I)^{P_2}, \hPi_J \rightarrow  (\hPi_J)^{P_2}$ also squares to itself, since 
\be
\label{transfer}
((\hX^I)^{P_2})^{P_2} = (\hX^I)^{P_2} - {\tilde P}_1 (\hX^I)^{P_2} {\tilde P}_1 = (\hX^I)^{P_2}
\ee
where we used the property ${\tilde P}_1^2={\tilde P}_1$.
This transformation is  therefore also a projection acting on the matrix operators $\hX^i, \hPi_J$. However, the transformation  does not act by conjugation, unlike $P_1$ for version 1).

The  notation we have adopted referring to the gauge invariant operators obtained in both cases, as ${\hat \cO}^{P_1}, {\hat \cO}^{P_2}$, allows for some simplification in the following discussion. We will often  refer to the operators obtained in both versions as   ${\hat O}^P$  without specifying which of the two cases $P_1, P_2$ we have in mind; where needed we will of course provide this clarification. 

In the subsequent discussion we will  also often denote the sub-algebra associated with a bulk region ${\cal R}$ which is obtained after projection, in either of the two versions as described above,  as ${\cal A}_R$. 

Let us end this subsection with two comments. First,  in general,  while passing from the constraint in terms of bulk coordinates, $f(x^I)$, eq.(\ref{condo}),
 to a constraint in terms of matrix operators, $f(\hX^I)$,  which appears  in the target space constraint,  we will encounter ordering ambiguities as was mentioned above. 
Note that in the matrix quantum mechanics, at any instant of time, the different matrix elements of the matrix operators commute, $[\hX^I_{ij},\hX^J_{kl}]=0$.
 However there are still matrix ordering ambiguities which are present since as matrices $[X^I,X^J]\ne 0$. 
 
As we will soon discuss, the matrix model we are dealing with is formulated in terms of matrix operators $X^1, \cdots X^9$, 
which correspond to the poincare coordinates in supergravity. For a linear constraint in the bulk involving these coordinates, where $f(x^I)=\sum _{I=1}^9 c_I x^I -a$, it is straightforward to obtain the operator constraint   $f(\hX^I)$ to be the corresponding function involving the matrix operators,  $f(\hX^i)=  \sum _{I=1}^9 c_I \hX^I-a$.
 For some of the non-linear constraints also there is a natural way  to find the corresponding  operator constraints,   for example 
$f(x^I)=\sum_{I=1}^9 c_I (x^I)^2 - a$, is mapped in a straightforward manner to  $f(\hX^I)=\sum_{I=1}^9 c_I (\hX^I)^2 - a$. 
In fact, the last example  can be extended to more general constraints which involve terms containing sums of monomials of individual coordinates, i.e.,
\be
\label{exam}
f(x^I)=\sum_I c_I (x^I)^{p_I} - a. 
\ee
 These are  mapped to 
\be
\label{exams}
f(\hX^i) = \sum_I c_I (\hX^I)^{p_I}- a. 
\ee
However, more general non-linear constraints involving terms with multiple coordinates cannot be mapped to a matrix constraint unambiguously, e.g.
take the case when $f(x^I)=x^1x^2-a>0$, this could be mapped to  either to $\hX^1 \hX^2-a>0$  or $\hX^2 \hX^1-a>0$. Our discussion below will primarily focus on  cases like eq.(\ref{exam})  where the map to the target space constraint eq.(\ref{exams}) is   straightforward\footnote{It could be  that such  operator ordering ambiguities  give rise to differences in entanglement entropy which are subheading in $N$, we thank S. Minwalla for making this comment. }. 

Second, it is easy to see that  the operators contained in the subalgebra ${\cal A}_R$ for  both versions 1) and 2) do not include  the Hamiltonian of the system. 
Remaining in the temporal gauge, let us rescale the matrices in (\ref{three-one}) and their conjugate momenta as
\ben
\hX^I \rightarrow (g_s N)^{1/3} l_s \hX^I~~~~~~~~~~~~~~\hPi^I \rightarrow \frac{1}{ (g_s N)^{1/3} l_s} \hPi^I
\label{three-four}
\een
the hamiltonian becomes
\ben
H = \frac{(g_sN)^{1/3}}{2 l_s}{\rm Tr} \left[ \frac{1}{N} \sum_{I=1}^9 (\hPi_I)^2  + N \sum_{I\neq J = 1}^9 [ \hX^I, \hX^J]^2 \right] + {\rm fermions}
\label{three-five}
\een
Instead ${\cal A}_R$  contains the operator 
\ben
H^P= \frac{(g_sN)^{1/3}}{2 l_s}{\rm Tr} \left[ \frac{1}{N} \sum_J[ (\hPi_J)^P]^2 + N \sum_{IJ} [ (\hX^I)^P,  (\hX^J)^P]^2 \right] + {\rm fermions} 
\een
 which is different.

In this paper we will consider gauge theories which involve matter fields in the adjoint representation. The main examples are gauged quantum mechanics of a single matrix, a particular example of which is the dual description of two dimensional strings, and $Dp$ brane field theories. The $D0$ brane quantum mechanics  is a particularly important   example  relevant for our discussion.

\subsection{Implementing a non-linear target space constraint}\label{gaugeinv2.3}

As described above, the gauge invariant formulation of target space entanglement applies to any constraint characterized by a hermitian operator $f(\hX^I)$. In a practical calculation, however, one would need to fix a gauge which diagonalizes this constraint. To perform a concrete calculation, however, one needs to make a change of variables to a set of {\em independent} variables which includes the eigenvalues. This is straightforward for a linear constraint, but becomes complicated very soon when we consider nonlinear constraints. In this subsection we explain how to do this for a constraint
\ben
f(\hX^I) = \sum_{I=1}^9 (\hX^I)^2 \equiv \hR^2
\label{9-1}
\een
The details of the procedure are given in the Appendix \ref{appone}.
What we need is a "polar" decomposition for matrices. 

\subsubsection{Two matrices}

Let us begin with the simplest case of two matrices $\hX^1$ and $\hX^2$. In the following all the matrices are operators in the Hilbert space unless stated otherwise.
We want to write this pair in terms of one hermitian matrix $\hR$ and a unitary matrix $\hQ$, where
\ben
\hR^2 = (\hX^1)^2 + (\hX^2)^2
\label{9-4a}
\een
Define the complex matrix
\ben
\hZ = \hX^1 + i \hX^2
\label{9-2}
\een
Then it follows that
\ben
2\hR^2 = \hZ \hZ^\dagger + \hZ^\dagger \hZ
\label{9-3}
\een
Now consider a singular value decomposition 
\ben
\hZ = \hV \hs \hW^\dagger
\label{9-4}
\een
where $\hV, \hW$ are unitary matrices and $\hs$ is a diagonal matrix. Using (\ref{9-3}) we then get
\begin{align}
2(\hR^2)_{ij}& = (\hV \hs^2 \hV^\dagger + \hW \hs^2 \hW^\dagger)_{ij} \nonumber \\
& = [ \hV^\star \otimes \hV + \hW^\star \otimes \hW ]_{ij,kl} (\hs^2)_{kl}
\label{9-5}
\end{align}
where 
\ben
[ \hV^\star \otimes \hV + \hW^\star \otimes \hW ]_{ij,kl} \equiv 
\hV_{ik} \hV^\star_{jl} + \hW_{ik} \hW^\star_{jl}
\label{9-6}
\een
It is shown in the Appendix \ref{apptwo} that the direct product matrix appearing in (\ref{9-5}) is invertible in the sense
\ben
[ ( \hV^\star \otimes \hV + \hW^\star \otimes \hW )^{-1} ]_{mn,ij}
[(\hV^\star \otimes \hV + \hW^\star \otimes \hW)]_{ij,kl} = \delta_{mk} \delta_{nl}
\label{9-7}
\een
An explicit expression for the inverse is
\ben
[ ( \hV^\star \otimes \hV + \hW^\star \otimes \hW)^{-1} ]_{kl,rs}
= \sum_{n=0}^\infty (-1)^n [\hV^\dagger (\hQ^\dagger)^n ]_{kr} [\hV^T (\hQ^T)^n]_{ls}
\label{9-8}
\een
where we have defined the unitary matrix $\hQ$
\ben
\hQ \equiv V W^\dagger
\label{9-9}
\een
We can now invert (\ref{9-5}) to write
\ben
\hs^2 = 2 \hV^\dagger \left[ \sum_{n=0}^\infty (-1)^n (\hQ^\dagger)^n \hR^2 \hQ^n \right] \hV
\label{9-10}
\een
In Appendix \ref{apptwo} we show that the matrix which appears in the square bracket in (\ref{9-10}) is positive semi-definite, so that we can take the square root of this equation. 
Substituting this in (\ref{9-4}) and using the definition of $\hQ$ in (\ref{9-9}) we finally get 
\ben
\hZ = ({\mathfrak{L}}_{\hQ} \hR) \hQ
\label{9-12}
\een
where we have defined the hermitian matrix ${\mathfrak{L}}_{\hV} \hM$,
\ben
{\mathfrak{L}}_{\hV} \hM \equiv  {\sqrt{2}} \left[ \sum_{n=0}^\infty (-1)^n (\hV^\dagger)^n \hM^2 \hV^n \right]^{1/2}
\label{9-13}
\een
where $\hV$ is unitary and $\hM$ is hermitian. This satisfies the equation
\ben
\hV^\dagger ({\mathfrak{L}}_{\hV} \hM)^2 \hV + ({\mathfrak{L}}_{\hV} \hM)^2 = 2 \hM^2
\label{9-13a}
\een
The matrices $\hX^1,\hX^2$ can be then expressed as 
\bea
\hX^1	& = & \frac{1}{2} \left[ ({\mathfrak{L}}_{\hQ} \hR) \hQ + \hQ^\dagger ({\mathfrak{L}}_{\hQ} \hR) \right] \nonumber \\
\hX^2 & = & \frac{1}{2i} \left[ ({\mathfrak{L}}_{\hQ} \hR) \hQ - \hQ^\dagger ({\mathfrak{L}}_{\hQ} \hR) \right] 
\label{9-14}
\eea
This is the matrix analog of a polar decomposition of cartesian coordinates in two dimensions $x^1 = r \cos \phi, x^2 = r \sin \phi$. For matrices we have the correspondence
\ben
r e^{i\phi} \rightarrow ({\mathfrak{L}}_{\hQ} \hR) \hQ 
\label{9-14a}
\een

To construct the relevant subalgebra of operators we then need to use the projector (\ref{proms}) with  $f(X^I) = \hR^2$, replace $\hX^I, I =1,2$ using (\ref{9-14}) by their projected versions (\ref{changever1}) and express the $\hX^I$ in terms of $\hQ$ and $\hR$ using (\ref{9-14}). 

An appropriate gauge-fixed version can be obtained by diagonalizing the matrix $\hR$,
\ben
\hR \rightarrow {\rm diag} [ \hr_1,\hr_2,\cdots \hr_N ]
\label{9-15}
\een
We can then proceed to work in a Hilbert space basis which are eigenstates of $\hr_i$ and the $\hQ_{ij}$ with eigenvalues $r_i,Q_{ij}$. The measure of integration then becomes
\ben
[dX^1 dX^2] = \mathbb{J} (r_i, Q_{ij}) \prod_i dr_i \prod_{ij} [dQ_{ij}]
\label{9-16}
\een
The jacobian $\mathbb{J} (r_i, Q_{ij})$ can be obtained in principle by using the explicit expressions (\ref{9-14}). However this is rather complicated, and we have not been able to obtain compact expressions for this. To proceed further, it will be convenient to write the unitary matrix $Q$ in terms of a unitary matrix $U$ and a set of angles $\phi_i$
\ben
Q = U e^{i\Phi} U^\dagger,~~~~~~\Phi = {\rm diag} (\phi_1, \phi_2, \cdots \phi_N)
\label{9-17}
\een
Defining
\ben
dS \equiv U^\dagger dU
\label{9-18}
\een
the line element then becomes
\ben
{\rm tr}(dQ dQ^\dagger) = \sum_i d\phi_i^2 + 8 \sum_{i < j} \sin^2 (\frac{\phi_i - \phi_j}{2}) dS_{ij} dS^\star_{ij}
\label{9-19}
\een
which leads to the expression
\ben
[dX^1 dX^2] =  \mathbb{J} (r_i, \phi_i, S_{ij}) \prod_i dr_i \prod_i d\phi_i  
\prod_{i < j} [ 4   \sin^2 (\frac{\phi_i - \phi_j}{2}) dS_{ij} dS^\star_{ij} ]
\label{9-20}
\een
As in the case of simple linear constraints the projector leads to restriction of the integration range of $r_i$. 

The above construction is inspired  by the work \cite{masuku} where the complex matrix $\hZ$ was written as $\hZ = {\tilde{R}}{\tilde{U}}$ where ${\tilde{R}}$ is a hermitian matrix operator and ${\tilde{U}}$ is a unitary operator. However in this decomposition ${\tilde{R}}^2 = (\hX^1)^2 + (\hX^2)^2 + i [ \hX^1 , \hX^2 ]$ rather than (\ref{9-4a}).

\subsubsection{Multiple Matrices}

The above polar decomposition can be extended to an arbitary number of matrices $\hX^I$. To illustrate the procedure let us first consider the case of three matrices $\hX^1,\hX^2,\hX^3$. The idea is to mimick the procedure to obtain spherical polar coordinates $(r,\theta,\phi)$ from usual cartesian coordinates $(x^1,x^2,x^3)$,
\ben
x^1 = r\cos \phi_1 \cos \phi_2,~~~x^2 = r\cos \phi_1 \sin \phi_2,~~~~x^3 = r \sin \phi_1
\label{11-1}
\een
We want to make a change of variables from hermitian matrices $\hX^I, I = 1,2,3$ to a hermitian matrix $\hR$ and two unitary matrices $\hQ_1,\hQ_2$. Here the matrix $\hQ_1$ generalizes $e^{i\theta}$, while $\hQ_2$ generalizes $e^{i\phi}$. From (\ref{9-14a}) the necessary replacements are 
\bea
r e^{i\phi_1 } & \rightarrow & ({\mathfrak{L}}_{\hQ_1} \hR) \hQ_1 \nonumber \\
r \cos \phi_1 e^{i\phi_2} & \rightarrow & \frac{1}{2} \left[{\mathfrak{L}}_{\hQ_2} \left(
 ({\mathfrak{L}}_{\hQ_1} \hR) \hQ_1+ \hQ_1^\dagger ({\mathfrak{L}}_{\hQ_1} \hR) \right)\right] \hQ_2
\label{11-2}
\eea
This leads to the final expressions
\bea
\hX^1 & = & \frac{1}{4} \left[{\mathfrak{L}}_{\hQ_2} \left(
 ({\mathfrak{L}}_{\hQ_1} \hR) \hQ_1+ \hQ_1^\dagger ({\mathfrak{L}}_{\hQ_1} \hR) \right)\right] \hQ_2 + \frac{1}{4} Q_2^\dagger \left[{\mathfrak{L}}_{\hQ_2} \left(
 ({\mathfrak{L}}_{\hQ_1} \hR) \hQ_1+ \hQ_1^\dagger ({\mathfrak{L}}_{\hQ_1} \hR) \right)\right] \nonumber \\
\hX^2 & = & \frac{1}{4i} \left[{\mathfrak{L}}_{\hQ_2} \left(
 ({\mathfrak{L}}_{\hQ_1} \hR) \hQ_1+ \hQ_1^\dagger ({\mathfrak{L}}_{\hQ_1} \hR) \right)\right] \hQ_2 - \frac{1}{4i} Q_2^\dagger \left[{\mathfrak{L}}_{\hQ_2} \left(
 ({\mathfrak{L}}_{\hQ_1} \hR) \hQ_1+ \hQ_1^\dagger ({\mathfrak{L}}_{\hQ_1} \hR) \right)\right] \nonumber \\
\hX^3 & = & \frac{1}{2i} \left(
 ({\mathfrak{L}}_{\hQ_1} \hR) \hQ_1- \hQ_1^\dagger ({\mathfrak{L}}_{\hQ_1} \hR) \right)
\label{11-3}
\eea
Using (\ref{9-13a}) repeatedly it is easy to see that
\ben
(\hX^1)^2 + (\hX^2)^2 + (\hX^3)^2 = \hR^2
\label{11-4}
\een
However the domain of the unitary matrices need to be restricted. This is because in (\ref{11-1}) one has $ -\pi/2 < \phi_1 < \pi/2$ while $ -\pi < \phi_2 < \pi $. To obtain the corresponding restriction on the domain of the unitary matrices $Q_1,Q_2$, we resort to the decomposition in (\ref{9-17}) for each of these matrices.
\ben
Q_A= U_A e^{i\Phi_A} U_A^\dagger,~~~~~~\Phi_A = {\rm diag} [(\phi_A)_1, \cdots
(\phi_A)_N]~~~~~~A = 1,2
\label{11-5}
\een
It is shown in the Appendix \ref{apptwo} that the requirement that the eigenvalues of $X^1 \cdots X^3$ should cover $\mathbb{R}^3$ once is equivalent to the requirement that 
\ben
 -\pi/2 < (\phi_1)_i < \pi/2~~~~~  -\pi < (\phi_2)_i < \pi
\label{11-6}
\een
The measure of integration is now
\ben
[dX^1 dX^2 dX^3] =  \mathbb{J} (r_i, (\phi_A)_i, (S_A)_{ij}) \prod_i dr_i  \prod_{A=1}^2 \left[ \prod_i d(\phi_A)_i  
\prod_{i < j} [ 4   \sin^2 (\frac{(\phi_A)_i - (\phi_A)_j}{2}) d(S_A)_{ij} d(S_A^\star)_{ij} ] \right]
\label{11-7}
\een
where we have defined, in analogy with (\ref{9-18})
\ben
dS_A \equiv U_A^\dagger dU_A
\een

It is now clear that this construction generalizes to arbitrary number of matrices $\hX^I, I = 1 \cdots D$. Once again we start with the polar coordinats of $\mathbb{R}^D$ and generalize to matrix polar decompositions which generalize (\ref{11-2}). Now we have a single hermitian matrix $\hR$ and $(D-1)$ unitary matrices $Q_A, A = 1,\cdots (D-1)$. Once each of the $Q_A$'s are decomposed as in (\ref{11-5}) we have the domains
\bea
-\pi/2 & < & (\phi_A)_i < \pi/2~~~~~~~A = 1 \cdots D-2 \nonumber \\
-\pi & < & (\phi_{(D-1)})_i < \pi
\label{angle}
\eea
The integration measure is as in (\ref{11-7}) with $A= 1 \cdots (D-1)$.

As discussed below we would like to identify the entanglement entropy associated with a constraint which restricts the eigenvalues $r_i$ to be in some range, e.g. $r_i > a$ in the dual supergravity background, at least for sufficiently large values of $a$.

\subsection{Dp Branes}

The above considerations generalize for Dp branes for $p < 3$. Now the matrices are scalar fields $X^I (\xi)$ and gauge fields $A_\mu (\xi)$. The target space restrictions are now on entire functions. The projector can be written in terms of a functional integral
\ben
\hP_A = \int {\cal D} x(\xi)~\prod_{\xi} \delta (x(\xi) - F[\hX^I(\xi)])
\label{8-1}
\een
where the functional $F[\hX^I(\xi)]$ needs to be chosen appropriately. For example, for a "planar" constraint we have $F[X^I(\xi)] = \hX^1(\xi)$. Once again one can choose a gauge which is tailored to the constraint, e.g. for the planar constraint we can pick a gauge where $\hX^1(\xi)$ is diagonal in matrix space. The discussion above can be now repeated and it follows that in this gauge one recovers the results of \cite{Das:2020jhy}.

\section{Target Space Entanglement as Bulk Entanglement}\label{tsebe}

Some further remarks are called for at this stage.  The purpose of our investigation is  to try and obtain  a precise version of bulk entanglement by mapping a bulk region to the target space in the boundary theory. Such an investigation is of course closely tied to understanding how approximate bulk locality arises in the boundary theory. 
In section \ref{gaugeinv} we showed how a sub-algebra of observables can be associated with the corresponding target space constraint. 
In this subsection we will discuss in what sense this subalgebra can probe supergravity fields in the region ${\cal R}$.

 Our main reasoning comes from the bulk meaning of the Coulomb branch of the gauge theory. 
Let us start with the system being in the vacuum. For  this case  the map we are using between the bulk and target space  is in agreement with what is well known about the system  in the moduli space approximation. A single $D0$ brane moving in the supergravity bulk dual to the D0 brane ground state experiences a velocity dependent potential which depends on its bulk location.This may be calculated by considering the DBI + Chern Simons action in the non-trivial background of a large number of D0 branes (\ref{ten-2}). This calculation is summarized in Appendix \ref{appthree}. For small velocities, the coefficient of $v^2$ is a constant, the next term goes like $v^4/r^7$, where $v$ is its velocity and $r$ the distance from the origin}. Exactly the same potential follows from  the gauge theory if the D brane location $(x^1, \cdots x^9)$ is mapped to a point  in the Coulomb branch with one  non-zero eigenvalue  for the matrices $X^1, \cdots X^9$.  For a D0 brane brane displaced along the $x^1$ direction and lying at $x_0^1$ the matrix $X^1$ has one non-zero eigenvalue, 
\begin{equation}
\label{singleprobe}
X^1 =
\begin{pmatrix}
{\bf 0}_{(N-1) \times (N-1)} & 0_{ 1 \times (N-1)} \\
0_{(N-1) \times 1} & x_0^1 =r + vt
\end{pmatrix}
\end{equation}
corresponding to $SU(N) \rightarrow SU(N-1) \times U(1)$. This represents  stripping off a  single D0 brane from a bunch of $N-1$ D0 branes which form a bound state.  Non-renormalization theorems ensure  the agreement, once this identification is made between the  bulk  and moduli space, see e.g., \cite{rseven,twoloop,polbecker} and references therein. 

At next order  a two loop calculation in D0 brane quantum mechanics yields a term which behaves as $v^6/r^{14}$. This term can be also reproduced  from the DBI+CS action as discussed in Appendix \ref{appthree} \footnote{A   bulk calculation can also be done in M theory with a compact null direction where $v$ is the relative velocity between two eleven dimensional gravitons with momenta $N_1/(g_sl_s)$ and $N_2/g_sl_s$ in the M theory direction for $N_2 \ll N_1$. The effect of the graviton with momentum $N_1/(g_sl_s)$ is to produce an Aichelburg-Sexl metric and the other graviton is considered as a probe in this background. The Aichelburg-Sexl metric results from an infinite boost of a 11 dimensional Schwarzschild black hole. We are interested in the extremal D0 brane background in 10 dimensions, which is obtainable from 11 dimensions by infinitely boosting a 11 dimensional black string along the string direction. While this looks like a different limiting procedure, the expansion of the DBI+CS action (see Appendix \ref{appthree}) in the latter background is exactly identical to the particle action considered in \cite{polbecker}.}

We can think of the calculations in the gauge theory as a computation of the effective action for appropriate gauge invariant operators.
In the example above, eq.(\ref{singleprobe}) we  calculate the effective action for the operators, 
\be
\label{set}
Tr(X^1), Tr(X^1)^2, \cdots Tr(X^1)^{N-1}
\ee
 and then evaluate this effective action by setting $Tr(X^1)^p=(x_0^1)^p$. The resulting  value of the effective action as  a function of $x_0^1$ then gives the effective potential for the probe $D0$ brane. 

There is, conceptually speaking,   another way to arrive at the same result in the gauge theory. Consider a region ${\cal R}$ which includes the location of the probe brane, i.e., 
\be
\label{inure}
(x_0^1, {\vec 0})\in {\cal R}
\ee
 As discussed in Section \ref{gaugeinv2.2}, given some region ${\cal R}$ we can define a target space constraint and an associated projector, leading to a sub-algebra of operators
${\cal A}_{\cal R}$. This algebra also consists of  gauge invariant operators  and we can 
also obtain an effective action for operators in ${\cal A}_{R}$, i.e. obtain the Legendre transformation of the generating function for  operators in ${\cal A}_R$. In this effective action we now set $Tr((X^1)^P))^m=(x_0^1)^m$, where the projector $P=P_1 \  {\rm or}  \ P_2,$  depending on whether we are considering version 1) or 2) of our proposal. The result, as a function of $x_0^1$ will then   agree with the effective action for the set of operators eq.(\ref{set}), and therefore will correctly give the potential experienced by the probe brane in the bulk,  as long as eq.(\ref{inure}) is met. This is manifestly clear if we  think of calculating the effective action using the  background field method in the gauge where $X^1$ is diagonal with background value given by eq.(\ref{singleprobe}), since we will then be doing the same calculation in the two cases. 

The force on the D0 brane in the bulk in the example above  can be calculated by using a DBI +CS action which is sensitive to the local values of the metric, the 10 dimensional $U(1)$  RR gauge field  and the dilaton. If the state is changed from the vacuum to some other coherent state $|s>$ which leads to a different background value of the metric and other bulk fields, we expect that the force that the probe D0 brane experiences can continue to be obtained in this way and will be sensitive to the local values of these bulk fields. For concreteness consider the state $|s>$ to  contain a gravity wave. Now, one way to obtain the local value of the gravitational field can be to measure how the potential for a probe brane at the location changes due to the presence of this gravity wave. This should yield the same result as other methods which may not involve a probe brane.

In the gauge theory  we expect that the potential for the probe brane continues to correspond to the effective action computed for suitable values of operators, i.e. the set eq.(\ref{set}), now in the state $|s>$ of the type we are considering, and we also expect that this effective action is correctly obtained from the effective action for  operators in ${\cal A}_{\cal R}$ as in the discussion above for the vacuum state. In this way we see that one expects to be able to obtain the one point function for the graviton, and some other supergravity modes,  from operators contained in ${\cal A}_R$.

This reasoning above is in fact  at the heart of our proposal for identifying the bulk and boundary target space regions and also identifying the algebra ${\cal A}_{\cal R}$ in the manner we have done.  In the sector where $k$ branes are present in ${\cal R}$  we keep the $k\times k$ block $M_{ab}$ for all matrix operators but not the complementary  $(N-k)\times (N-k)$ block which correspond to branes that are not present. In version 1) of our proposal we also keep the off diagonal blocks $M_{a\alpha}, M_{\alpha,a}$. The resulting algebra  of observables allows one to describe all measurements done on branes present in ${\cal R}$ and this should then  be sufficient to also  detect low-energy  supergravity excitations in ${\cal R}$. 

It is worth noting in this context  that there is additional  evidence, going beyond the moduli space approximation, that the map  between the bulk and target space  we are using continues to work. For example, one can consider two stacks  of D0 branes one at the origin and the other displaced from it and excite  open strings  within  branes in each set. This changes the potential between the two stacks, but one still finds agreement in the bulk and in the gauge theory for the resulting two-body interactions, after appropriately identifying 
operators in the gauge theory  with their counterpart currents   in the bulk, \cite{rseven}. This suggests that the algebra ${\cal A}_{\cal R}$, which retains the appropriate operators for all the superselection sectors where different number of branes $k=0,1, \cdots N$ are present in the region of interest, should suffice for describing the results of all measurements made with local supergravity operators in ${\cal R}$. 
 
Our intuition based on the above reasoning,  can be extended to a given {\em configuration} in the bound state wavefunction for $N$ D0 branes in the gauge where the constraint function is diagonal. Consider a configuration where $k$ of these eigenvalues are in the region of interest ${\cal R}$. The degrees of freedom $M_{a\alpha}, M_{\alpha a}$ correspond to  excitations in the bulk going  between ${\cal R} $ and ${\cal R}^c$ or vice versa. If the state $|s>$ has some supergravity modes excited with support deep inside ${\cal R}$, by which we mean the excitations are localised many string lengths away from the boundary of ${\cal R}$, then neither the $M_{\alpha\beta}$ nor the $M_{a \alpha}, M_{\alpha a}$ degrees of freedom will be excited in $|s>$. The $M_{\alpha \beta}$ degrees will not be excited because the excitations in $|s>$ are   localised in ${\cal R}$. The $M_{a \alpha}, M_{\alpha a}$ degrees will also not be excited because they would correspond to open strings which would have to stretch across the boundary across many string lengths and would therefore be very heavy. 

As a result, one might expect that the full change in expectation values of any single trace operator  ${\hat O}$ made out of a string of $X^I,\Pi_J$'s  schematically depicted in eq.(\ref{ctrace}) will be obtained to good approximation by the corresponding operator obtained after projection,  ${\hat O}^P$, eq.(\ref{opproj}), eq.(\ref{ver2op}) . 
Notice that at this level of admittedly  imprecise arguments we cannot distinguish between version 1) and 2) of our proposal. Both contain the gauge invariant degrees of freedom coming from $M_{ab}$. In version 1) there are extra degrees of freedom coming from  $M_{a\alpha}, M_{\alpha a}$  as well but as per the intuitive argument above they might not play an important role anyways. 

However, there are reasons to believe that the sub-algebra we are considering  will not provide all details of bulk fields as defined e.g. by the
  BDHM-HKLL map. In the low energy regime such a bulk field operator $\phi(r, \theta_i, t)$ is defined by
\be
   \label{expolo}
   \phi(r, \theta_i, t)= \sum_{l_i, \omega_n} [  {\hat O}_{l_i, \omega_n} f_{l_i, \omega_n}(r) Y_{l_i}(\theta_i) e^{-i \omega_n t} + cc]
  \ee
where ${\hat O}_{l_i,\omega_n}$ are Fourier modes with frequency $\omega_n$  obtained from the  time dependent operators ${\hat O}_{l_i}(t)$ (we are being schematic here, $\omega_n$ need not be discrete). 
Consider such a bulk operator with $(r,\theta_i)\in {\cal R}$.
The expectation value of $\phi$ can  be  obtained if we know  the expectation value of  ${\hat O}_{l_i}(t)$ for all $l_i$ and all times $t$. Now we can regard ${\hat O}_{l_i}(t)$ as an operator acting on the Hilbert space of states  at $t=0$. Its  expectation values can therefore be obtained, in principle, if the expectation values of all operators are known at time $t=0$. In this way we see that the expectation value of $\phi(r,\theta_i,0) $ in any state can be obtained once  the expectation values of all operators in the  corresponding state in the boundary theory are  known at  $t=0$ . One such bulk operator is  the metric itself, for which the corresponding gauge theory operator is the energy momentum tensor.

Consider now a region ${\cal R}$ which is a small annular region near the boundary, 
\be
\label{annreg}
r_B^2-\delta<r^2<r_B^2, 
\ee
where $r_B$ is the boundary value of the radial variable $r^2=\sum_{i}(x^i)^2$, and $\delta$ is small. 
It has been argued in \cite{suvrat}, \cite{suvrat2}, that measurements carried out by observers in this region will allow detailed information about  the state  in the bulk to be obtained. It is crucial in these arguments that the observers close to the boundary have access to the full Hamiltonian of the  system. In fact, having  access to   the Hamiltonian alone enables observers  in ${\cal R}$   to  reconstruct the full density matrix of the vacuum, $|0><0|$. As a result any algebra, which includes all operators corresponding to measurements  bulk observers in the region eq.(\ref{annreg}) can make, in particular which includes  the Hamiltonian,  would   have  a vanishing  entanglement entropy for  the vacuum state\footnote{We are grateful to Suvrat Raju for explaining this point to us.}. 
The subalgebra we are associating with the region ${\cal R}$ are the projected versions of the gauge theory operators, ${\hat {\cal O}}_{l_i\omega_n}$, which appear in (\ref{expolo}). Thus, the bulk operators defined in eq.(\ref{expolo}), with the restriction that $(r,\theta_i)\in {\cal R}$, are not contained in this subalgebra. In particular the  Hamiltonian is not an element of this subalgebra, only its projected version is. We expect that this imposes significant restrictions on the amount of information which can be obtained for the bulk operators.  In this sense our  sub-algebra would only  capture the notion of a local bulk region in an approximate sense.  A more detailed investigation of how significant these restrictions are is left for the future.

Clearly, one would  like a deeper understanding of the various issues discussed in this section. In particular, one would like a better understanding of how much information about the bulk region of information  can actually be obtained from  the sub-algebras we propose, and also whether there  are refinements to our basic proposal, including an improved map between the constraint in the bulk eq.(\ref{condo}) and the corresponding one in the boundary eq.(\ref{cond2}), that are needed. 
Since the issues at had  are   closely  tied to how approximate bulk locality arises, as was mentioned above at the outset, 
it is unlikely though  that we can make much progress   through analytic methods  alone . Numerical calculations  hold considerable promise in this regard. Roughly speaking one wants to show that the change in the wave function which correspond to changes in some  bulk region ${\cal R}$, arises mainly in the target space region associated with ${\cal R}$ and not its complement. This should also then shed light on which  operators  would be needed to determine  this change in the boundary theory and whether a sub-algebra along the lines proposed here would suffice.

\section{Target Space Entanglement and Bekenstein Bound}\label{tsebb}

In the 't Hooft limit the usual 't Hooft limit $g_s \rightarrow 0, N \rightarrow \infty$ with $(g_sN)$ held fixed, the bound state of $N$ Dp branes is dual to a ten dimensional geometry. Our proposal implies that target space constraints correspond to regions in the transverse space to these Dp branes in this geometry and the target space entanglement entropy defined above provides a notion of a bulk entanglement entropy associated with this region. In \cite{Das:2020jhy} it was conjectured that for Dp brane matrix field theories, the target space entanglement entropy saturates the Bekenstein bound for this entangling surface. In this section we recapitulate the result for D0 branes.

We will be mostly interested in the bound state of $N$ D0 branes which are slightly heated up to a temperature $T$. This is dual to the near-extremal black D0 brane geometry in supergravity. The string frame metric, dilaton and 1-form gauge fields are
\bea
ds_{string}^2  & =  & -H_0(r)^{-1/2} g(r) dt^2 + H_0(r)^{1/2}[\frac{dr^2}{g(r)}+r^2 d\Omega_8^2] \nonumber \\
e^{\phi}  & =  &g_s H_0(r)^{3/4}~~~~~~~~~~~~A_0 = -\frac{1}{2}(H_0^{-1}-1)
\label{ten-2}
\eea
where
\ben
g(r)  =  1- \left(\frac{r_H}{r} \right)^7 ~~~~H_0(r)  =  \frac{R^7}{r^7}, 
~~~~~~r^2  =  x_1^2 + \cdots x_9^2.
\label{ten-1}
\een
The horizon is at $r=r_H$. The Hawking temperature for this solution and the length scale $R$ are  given by
\ben
T  =  \frac{7}{4\pi R} \left(\frac{r_H}{R} \right)^{5/2} ~~~~~R^7  =  60 \pi^3 l_s^7(g_s N).
\label{eleven-1}
\een
The supergravity solution above is valid in the regime
\ben
g_s^{\frac{1}{3}} N^{1/7} < r < (g_s N)^{1/3} l_s~~~~~~~~T \frac{l_s}{(g_s N)^{1/3}} \ll 1
\label{three-14}
\een
In D0 brane quantum mechanics consider the simple linear constraint, e.g. $f(X^I) = X^1 - a_0$. According to the proposal of \cite{Das:2020jhy} the bulk region of interest is simply $x^1 > a$. The relationship between the dimensionless $a_0$ and the dimensionful $a$ can be read off from the rescaling (\ref{three-four})
\ben
a = a_0 (g_sN)^{\frac{1}{3}} l_s
\label{three-11}
\een
This reflects the fact that in this holographic correspondence the transverse distance becomes the energy scale of the D0 brane quantum mechanics which is $\Lambda = (g_sN)^{1/3}/l_s$. Likewise the temperature appearing in (\ref{eleven-1}) is related to a dimensionless temperature $T_0$ by $T = T_0 \Lambda$. 

In \cite{Das:2020jhy} it was conjectured that this target space entanglement saturates the Bekenstein bound
\ben
S (a,T) = \frac{A_a (T)}{4 G_N}
\label{three-16}
\een
where $A_a$ is the {\em Einstein Frame} area of the entangling surface $x^1 = a$ in the geometry (\ref{ten-2}) and $G_N = 8\pi^6 g_s^2 l_s^8$ is the ten dimensional Newton constant. 
The quantity $S(a,T)$ is actually divergent, the divergence coming from the large $r$ region. The large $r$ region is, however, beyond the regime of validity of supergravity: thus one may consider using a cutoff at $r=r_0$. However the difference $S(a,T) - S(a,T^\prime)$ is finite, 
\ben
S(a,T) - S(a,T^\prime) = B_0~ N^2  a_0^{-5/2}\left[ (T_0)^{14/5}-(T_0^\prime)^{14/5} \right]
\label{three-17}
\een
where $B_0$ is a number whose value is given in equation (29) of \cite{Das:2020jhy}.

Note that the expression (\ref{three-17}) the dimensionless quantities which characterize the state and the entangling region are those which are quantities which would appear in D0 brane quantum mechanics. The only other number which appears is $N$ : the answer is proportional to $N^2$. This is what one expects if our proposal is correct. In particular all factors of $g_s$ nicely cancel. The powers of $T_0$ and $d_0$ which appear in (\ref{three-17}) does not follow from general considerations of target space entanglement. If a numerical calculation yields these powers we will have a very non-trivial evidence for our proposal.

 Let us make one comment before ending this section. We have emphasised above that the discussion in this paper can be applied for constraints taking the general form, 
eq.(\ref{condo}). Instead of the linear constraint considered above suppose  we  take 
\be
\label{constantszz}
f(x^1)=\sum_{i=1}^9 ( x^i)^2>r_0^2 
\ee  where $r_0$ is some radius.  In this case the Beckenstein- Hawking entropy  which is  given by 
\be
\label{bhee}
S=\Omega_8 {R^{7/2} r_0^{9/2}\over 4 G_N}
\ee
 is a function of $r_0$ but  is independent of the temperature $T$. 
As per our proposal we  would like to equate this result with the entanglement entropy associated with the  target space  constraint 
\be
\label{construe}
\sum_{I=1}^9({\hat X}^I)^2>r_0^2.
\ee
However it does seem rather strange then  that the resulting entanglement entropy is independent of the temperature $T$. One reason could be that perhaps the map between a physical region in the bulk and the corresponding target space constraint  is more complicated at finite temperature, i.e. the RHS in eq. (\ref{constantszz}) and eq.(\ref{construe})  are not equal but instead  related by a temperature dependent function. This might also help explain why when we take $r_0=r_H$ in eq.(\ref{constantszz}), we get the entanglement entropy to be the full entropy in the boundary theory and not a different value due to the additional target space constraint eq.(\ref{construe}) being present. We leave a more detailed investigation of such temperature dependent effects for the future.

\section{Path Integral Expressions for Renyi Entropies}\label{piere}

As discussed above, numerical calculations should be able to prove or disprove our conjecture that the target space von Neumann entropy saturates the Bekenstein bound. Recently there has been impressive advances in numerical calculations for D0 branes \cite{Hanada:2016zxj}. These calculations use euclidean path integrals to calculate finite temperature partition functions as well as some correlation functions. In this section we develop euclidean path integral expressions for target space Renyi entropies which can be used directly for numerical calculations. These expressions are in the gauge fixed formalism, and we will develop them for planar constraints.

Consider the D0 brane theory at some finite temperature $T = 1/\beta$. The density operator is given by $\hrho_0 = {\rm exp} [-\beta H]$  where the hamiltonian $H$ is given by (\ref{three-five}). As in the previous sections, we will fix the $A_0 =0$ gauge, fix the time independent gauge transformations by diagonalizing one of the matrices $X^1$, and impose the remaining Weyl and $U(1)^N$ symmetries by explicitly summing over the corresponding transformations. The basis states are given by (2.40). In the following we will also ignore the fermions.

In the absence of any symmetrization the matrix elements of $\hrho$ can be written as a path integral as follows
\bea
\langle \lambda_i,X^L_{ij}|\hrho_0 |\lambda_i^\prime, (X^L_{ij})^\prime \rangle = \int_{\lambda_i(0) =\lambda_i^\prime}^{\lambda_i(\beta) =\lambda_i}
 \cD \lambda_i (\tau) \int_{X^L_{ij}(0) = (X^L_{ij})^\prime}^
{X^L_{ij}(\beta) = X^L_{ij}}
 \cD X^L_{ij} (\tau) ~{\rm exp}[-S_\beta] 
\label{10-1}
\eea
where the action $S_\beta$ is the euclidean action 
\ben
S = \frac{(g_sN)^{1/3} N }{2 l_s} \int_0^\beta d\tau ~{\rm Tr} \left[ \sum_{I=1}^9 (\partial_\tau X^I)^2  + \sum_{I\neq J = 1}^9 [ \tX^I, \tX^J]^2 \right] 
\label{10-2}
\een
Weyl and $U(1)^N$ symmetries are then imposed by explicitly summing over the transformations, leading to braided boundary conditions. However, since the action is symmetric under these transformations, we need to sum over transforms of the boundary conditions at one of the ends of the euclidean time interval. We therefore have
\bea
& &  _W\langle \lambda_i,X^L_{ij}|\hrho_0 |\lambda_i^\prime, (X^L_{ij})^\prime \rangle_W  = \rho_{tot} (\lambda_i,X^L_{ij};\lambda_i^\prime, (X^L_{ij})^\prime) \nonumber \\
& = &~\frac{1}{N !} \int \prod_{i=1}^N \frac{d\theta_i}{2\pi}\sum_{\sigma \in S(N)} (-)^\sigma 
\int_{\lambda_i(0) =\lambda_i^\prime}^{\lambda_i(\beta) =\lambda_{\sigma(i)}}
\cD \lambda_i (\tau) \int_{X^L_{ij}(0) = (X^L_{ij})^\prime}^
{X^L_{ij}(\beta) = (X^L_{ij})^W}
 \cD X^L_{ij} (\tau) ~{\rm exp}[-S_\beta] 
\label{10-3}
\eea
where we have introduced the notation
\ben
(X^L_{ij})^W \equiv X^L_{\sigma(i)\sigma(j)} e^{i(\theta_{\sigma(i)}-\theta_{\sigma(j)})}
\label{weylu1}
\een
which we will use in the following equations as well.

The construction for $N=2$ and with two matrices $X^1, X^2$ is illustrated in Figure \ref{fig:one}. Note that each term in the path integral is {\em not} a product of path integrals over $\lambda_i$ and $X^L_{ij}$ since the interaction term in the action couple them. These interactions are symbolically drawn as rectangular boxes to emphasize this. The figure is meant to illustrate the boundary conditions.

\begin{figure}[!h]
\centering
\includegraphics[width=4.5in]{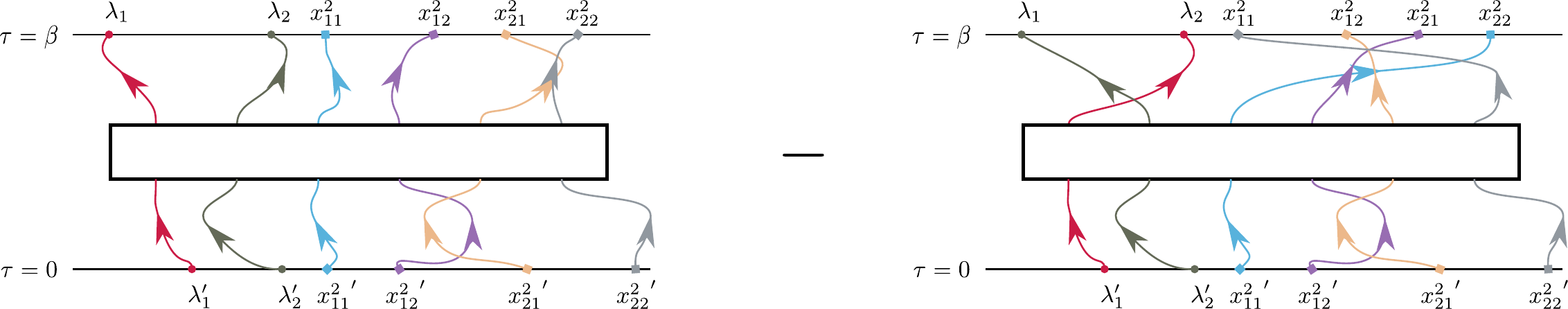}
\caption{Path Integral Representation for the thermal density matrix for a model of two $2 \times 2$ matrices $X^1$ and $X^2$ in the gauge where $X^1$ is diagonal with eigenvalues $\lambda_1$ and $\lambda_2$. The blobs represent arbitrary number of interactions between the paths. }
\centering
\label{fig:one}
\end{figure}

To obtain the reduced density matrix in some sector $(k,N-k)$ one needs to integrate over the appropriate set of boundary values. Consider some interval $A$ on the real line. As in the previous section we will split the matrix indices into two sets, $a,b = 1\cdots k$ and $\alpha, \beta =  k+1 \cdots N$ where the eigenvalues  $\lambda_a$ lie in $A$ while the remaining $\lambda_\alpha$ lie in the complement $\bA$. The boundary values of the matrix elements of $X^L$ with $L = 2 \cdots 9$ are not constrained in any fashion.
Then the expressions for the two proposals are given in (\ref{three-nine}) and (\ref{three-nine-1}). 

In terms of the paths in the path integral this means the following. Along a given path parametrized by $0 < \tau < \beta$, the  $\lambda_a(\tau)$ must begin and end in distinct points in the interval $A$. The eigenvalues $\lambda_\alpha (\tau)$ must begin and end at the {\em same point} in the complement $\bA$, and there is an integral over this point. It is important to note that apart from these restrictions the paths are free to wander around anywhere in the $\lambda$ space at intermediate times. 

In proposal (1), the boundary values of $X^L_{\alpha\beta}, X^L_{a\alpha}, X^L_{\alpha a}$ are the same at $\tau =0$ and $\tau = \beta$ and are integrated over, while the boundary values of $X^L_{ab}$ are different. This leads to the following expression for the reduced density matrix:
\begin{equation}
\begin{split}
& \tilde{\rho}_{k,N-k}^{(1)} (\lambda_a, X^L_{ab}; \lambda_a^{\prime}, (X^L_{ab})^{\prime}) \\
& =\frac{1}{N!} \binom {N}{k} \int_{\bar{A}} d \lambda_\alpha \int d X^{L}_{a\alpha} d X^{L}_{\alpha a} d X^{L}_{\alpha \gamma} \int \prod_{i=1}^N \frac{d\theta_i}{2\pi} 
\sum_{\sigma \in S(N)} (-)^\sigma 
\int_{\mathcal{A}_1}
\mathcal{D}\lambda_i (\tau) \int_{\mathcal{B}_1}
\mathcal{D} X^L_{ij} (\tau) ~{\rm exp}[-S_\beta]
\end{split}
\label{12-2}
\end{equation}
where the boundary conditions are denoted by
\begin{equation}
\mathcal{A}_1 = 
\begin{pmatrix}
\lambda_a(0) =\lambda_a^\prime , & \lambda_{\sigma(a)}(\beta) =\lambda_a \\
\lambda_\alpha(0) =\lambda_\alpha  ,& \lambda_{\sigma(\alpha)}(\beta) =\lambda_\alpha \end{pmatrix}
\label{12-4}
\end{equation}

\begin{equation}
\mathcal{B}_1 = 
\begin{pmatrix}
X^L_{ab}(0) = (X^L_{ab})^\prime , & X^L_{a\alpha}(0) = X^L_{a\alpha} , & X^L_{\alpha a}(0) = X^L_{\alpha a} , & X^L_{\alpha\gamma}(0) = X^L_{\alpha\gamma} \\
(X^L_{ab})^W (\beta) = X^L_{ab} , & (X^L_{a\alpha})^W (\beta)=X^L_{a\alpha} , & 
 (X^L_{\alpha a})^W (\beta)=X^L_{\alpha a} , & (X^L_{\alpha\gamma})^W (\beta) = X^L_{\alpha\gamma}
 \end{pmatrix}
\end{equation}

The Figures \ref{fig:two}-\ref{fig:four} show the paths for $N=2$ in the various sectors for our first proposal, drawn as paths on a cylinder which is cut across the region $A$.  In each sector there are two terms. In these figures we have represented only the boundary values of the eigenvalues of one of the matrices $X^1$. The other matrix elements are braided in the manner indicated in Figure \ref{fig:one}. As in Figure \ref{fig:one} these diagrams are illustrative of the boundary conditions: the rectangular boxes represent interactions between the variables along the paths.

\begin{figure}[!h]
\centering
\includegraphics[width=5.0in]{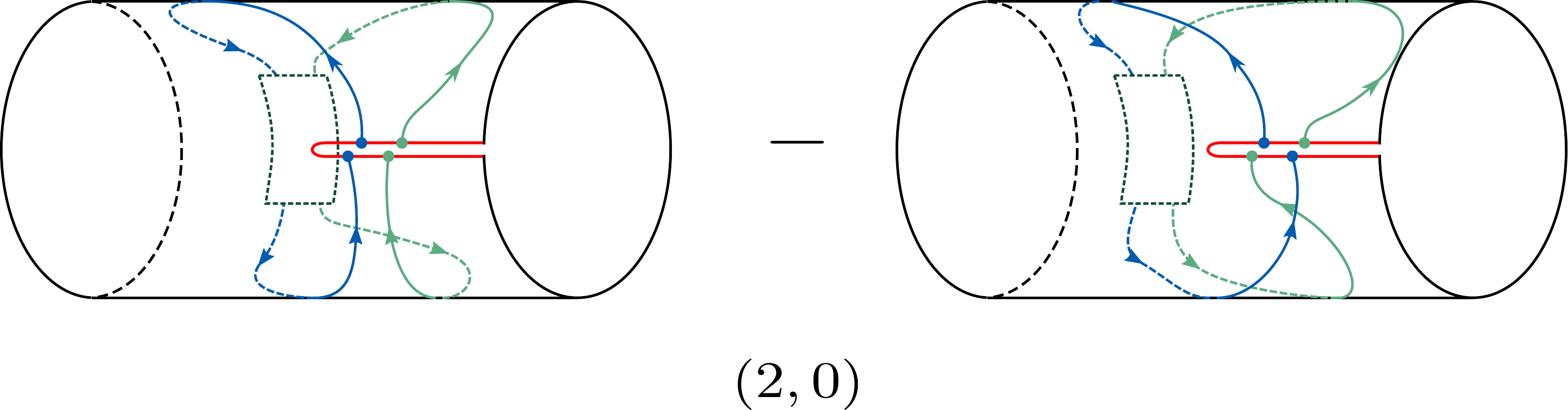}
\caption{Path Integral Representation for the reduced density matrix in the $(2,0)$ sector for a model of two $2 \times 2$ matrices $X^1$ and $X^2$ in the gauge where $X^1$ is diagonal. The red cut represents the region of interest $A$. We have shown the end-point values only for the eigenvalues of $X^1$.}
\centering
\label{fig:two}
\end{figure}

\begin{figure}[!h]
\centering
\includegraphics[width=5.0in]{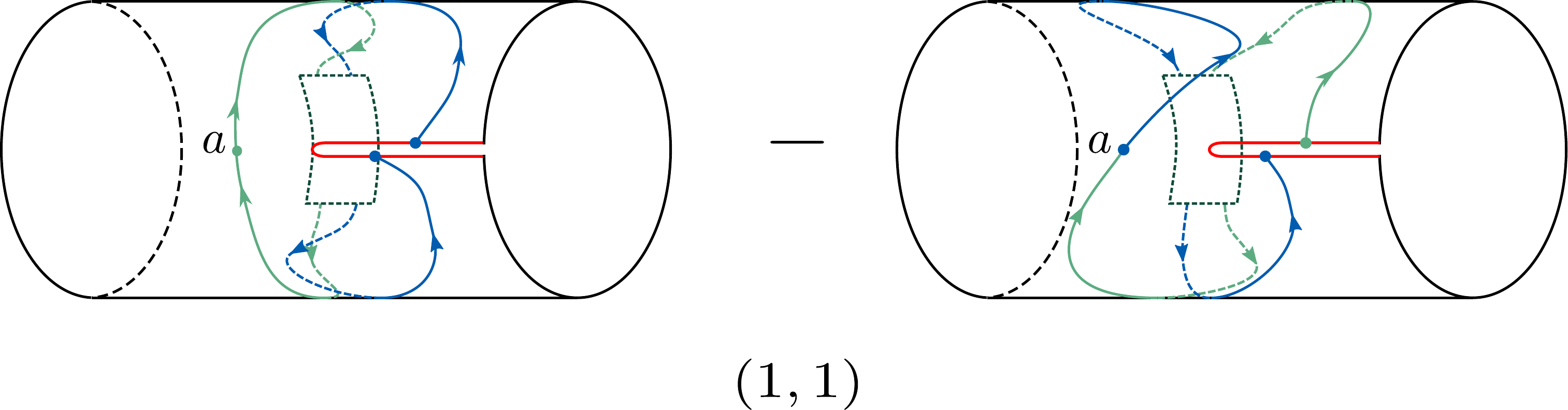}
\caption{Path Integral Representation for the reduced density matrix in the $(1,1)$ sector for a model of two $2 \times 2$ matrices $X^1$ and $X^2$ in the gauge where $X^1$ is diagonal. The red cut represents the region of interest $A$. We have shown the end-point values only for the eigenvalues of $X^1$.}
\centering
\label{fig:three}
\end{figure}

\begin{figure}[!h]
\centering
\includegraphics[width=5.0in]{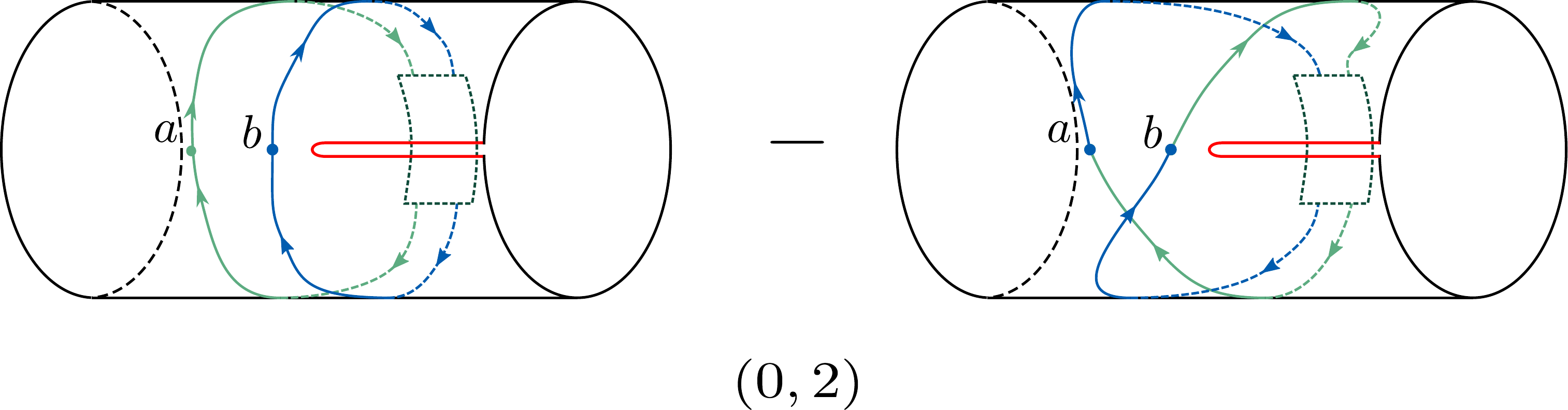}
\caption{Path Integral Representation for the reduced density matrix in the $(0,2)$ sector for a model of two $2 \times 2$ matrices $X^1$ and $X^2$ in the gauge where $X^1$ is diagonal. The red cut represents the region of interest $A$. There are no specified boundary values.}
\centering
\label{fig:four}
\end{figure}

In proposal (2), only the boundary values of $X^L_{\alpha\beta}$ are same and integrated over. 
\begin{equation}
\begin{split}
& \tilde{\rho}_{k,N-k}^{(2)} (\lambda_a, X^L_{ab}, X^L_{a\alpha}, X^L_{\alpha a}; \lambda_a^{\prime}, (X^L_{ab})^{\prime}, (X^L_{a\alpha})^{\prime}, (X^L_{\alpha a})^{\prime}) \\
& =\frac{1}{N!} \binom {N}{k} \int_{\bar{A}} d \lambda_\alpha \int d X^{L}_{\alpha \gamma} \int \prod_{i=1}^N \frac{d\theta_i}{2\pi}  \sum_{\sigma \in S(N)} (-)^\sigma 
\int_{\mathcal{A}}
\mathcal{D}\lambda_i (\tau) \int_{\mathcal{B}_2}
\mathcal{D} X^L_{ij} (\tau) ~{\rm exp}[-S_\beta]
\end{split}
\label{12-3}
\end{equation}
The boundary conditions for the $\lambda_i$ remain the same as in (\ref{12-4}) , while those for the $X^L_{ij}$ are denoted by

\begin{equation}
{\mathcal{B}}_2 = 
\begin{pmatrix}
X^L_{ab}(0) = (X^L_{ab})^\prime , & X^L_{a\alpha}(0) = (X^L)^\prime_{a\alpha} , & (X^L)_{\alpha a}(0) = (X^L)^\prime_{\alpha a} , & X^L_{\alpha\gamma}(0) = X^L_{\alpha\gamma} \\
(X^L_{ab})^W (\beta) = X^L_{ab} , & (X^L_{a\alpha})^W(\beta) =X^L_{a\alpha}  , & (X^L_{\alpha a})^W (\beta) =X^L_{\alpha a} , & (X^L_{\alpha\gamma})^W (\beta) = X^L_{\alpha\gamma}
 \end{pmatrix}
\end{equation}

The figures for paths for the second proposal can be drawn as in the earlier figures.

It is now straightforward to compute ${\rm tr} \trho^n_{k,N-k}$ by taking powers of these expressions and tracing. For example, ${\rm tr} \trho^2_{k,N-k}$ for Proposal (1) is
\begin{equation}
\begin{split}
& \operatorname{tr} (\tilde{\rho}_{k,N-k}^{(1)})^2 \\
=&\left[ \frac{1}{N!} \binom {N}{k} \right]^2  \int_{{A}} d \lambda_a d \lambda_a^\prime \int d X^{L}_{ab}d (X^{L}_{ab})^\prime 
 \times \int_{\bar{A}} d \lambda_\alpha \int d X^{L}_{a\alpha} d X^{L}_{\alpha a} d X^{L}_{\alpha\beta} 
 \int_{\bar{A}} d \lambda_\alpha^{\prime} \int d (X^{L}_{a \alpha})^\prime d (X^{L}_{\alpha a})^\prime d (X^{L}_{\alpha\beta})^\prime \\
&\times \int \prod_{i=1}^N d\theta_i \int \prod_{j=1}^N d\theta^\prime_j\sum_{\sigma, \sigma^\prime \in S(N)} (-)^{\sigma+\sigma^\prime}   \int_{\ccC_1}
\mathcal{D} \lambda_{i} (\tau) \int_{\ccC_2}
 \mathcal{D} X^L_{ij} (\tau) ~{\rm exp}[-S_\beta] 
\int_{\ccC_3}
\mathcal{D} \lambda_{i} (\tau) \int_{\ccC_4}
 \mathcal{D} X^L_{ij} (\tau) ~{\rm exp}[-S_\beta] \\
\end{split}
\end{equation}
where the periodicity conditions are
\ben
\ccC_1 =
\begin{pmatrix}
\lambda_a (0) =\lambda_a^\prime, &  \lambda_\alpha(0) =\lambda_\alpha \\
\lambda_{\sigma(a)}(\beta) =\lambda_a, &\lambda_{\sigma(\alpha)}(\beta) =\lambda_\alpha 
\end{pmatrix}
\een
\ben
\ccC_2 =
\begin{pmatrix}
X^L_{ab}(0) = (X^L_{ab})^\prime , & X^L_{a\alpha}(0) = X^L_{a\alpha},  & X^L_{\alpha a}(0) = X^L_{\alpha a}, & X^L_{\alpha \gamma}(0) = X^L_{\alpha \gamma} \\
 (X^L_{ab})^W(\beta) = X^L_{ab}, & (X^L_{a\alpha})^W (\beta) =X^L_{a \alpha} , & (X^L_{\alpha a})^W(\beta) =X^L_{\alpha a} , & (X^L_{\alpha\gamma})^W (\beta) = X^L_{\alpha\gamma} 
\end{pmatrix}
\een
\ben
\ccC_3 =
\begin{pmatrix}
\lambda_a(0) =\lambda_a, &  \lambda_\alpha(0) =\lambda_\alpha^\prime \\ \lambda_{\sigma^\prime(a)}(\beta) =\lambda_a^\prime, & \lambda_{\sigma^\prime(\alpha)}(\beta) =\lambda_\alpha^\prime
\end{pmatrix}
\een
\ben
\ccC_4 =
\begin{pmatrix}
X^L_{ab}(0) = X^L_{ab}, & X^L_{a\alpha}(0) = (X^L_{a\alpha})^\prime, & 
X^L_{\alpha a}(0) = (X^L_{\alpha a})^\prime, & X^L_{\alpha\gamma}(0) = (X^L_{\alpha\gamma})^\prime \\
(X^L_{ab})^{W^\prime}(\beta) = (X^L_{ab})^\prime , & (X^L_{a\alpha})^{W^\prime}(\beta) =(X^L_{a\alpha})^\prime ,
& (X^L_{\alpha a})^{W^\prime}(\beta) =(X^L_{\alpha a})^\prime , &  (X^L_{\alpha\gamma})^{W^\prime}(\beta) = (X^L_{\alpha\gamma})^\prime 
\end{pmatrix}
\een
where in an obvious extension of the notation of (\ref{weylu1})
\ben
(X^L_{ij})^{W^\prime} \equiv X^L_{\sigma^\prime (i)\sigma^\prime (j)} e^{i(\theta^\prime_{\sigma^\prime (i)}-\theta^\prime_{\sigma^\prime(j)})}
\een
The expression is invariant if we exchange $\sigma \leftrightarrow \sigma', \theta_i \leftrightarrow \theta_i^\prime$. The expression for ${\rm tr} \trho^n_{k,N-k}$ in Proposal (2) can be similarly written down. The Renyi entropies $S_n$ can be then computed using these expressions,
\ben
S_n = \frac{1}{n-1} \log \left[ \sum_{k=0}^N {\rm tr}_k \trho^n_{k,N-k} \right]
\een
These path integral expressions can be directly used in numerical calculations. It is difficult to take the $\beta \rightarrow \infty$ limit to recover a zero temperature answer. However it should be possible to compare the difference of the entropies at two different temperatures with the supergravity result.

\section{Conclusions} \label{cncl}
In this paper we have have proposed that for near -horizon $Dp$ brane backgrounds, target space entanglement in the boundary theory provides a precise version of bulk entanglement in the gravity dual. We have described how to obtain in a gauge invariant manner a sub -algebra related to a  target space constraint. A Von-Neumann entropy can be associated with this sub-algebra in the standard manner and this then gives the  target space entanglement entropy. Our  paper builds on \cite{Das:2020jhy} which dealt with linear constraints  in a gauge fixed formalism and we have provided here a general gauge invariant description of the target space entanglement. 

We  have  also  provided  some arguments here,  based on comparisons between the potential experienced by probe branes  moving in some region of the bulk ${\cal R}$ and the  effective potential in the boundary theory evaluated in the corresponding region of moduli space, to motivate  why  the sub-algebra of operators ${\cal A}_{\cal R}$  that we identify  is  sufficient to  describe some local bulk measurements  on gravitons and other supergravity modes, in addition to D branes, carried out by  observers in ${\cal R}$. 

We should emphasise that our arguments are not {\em completely} precise. One reflection of this is that  there are in fact two versions of our proposal, which give rise to two different sub-algebras related to a target space constraint, and we cannot distinguish between them at our current level of understanding. 

One source of  imprecision  in our proposal could be  that the  map between the bulk  and target space constraints,  eq.(\ref{condo})  and eq.(\ref{cond2}), is more complicated than we have assumed. This could happen  due  to operator ordering ambiguities or for excited states, including at finite temperature,  where some  or all of the supersymmetries are broken.  The target space function $f({\hat X}^i)$ which appears in eq.(\ref{cond2}) in such situations could be more non-trivially related to its bulk counterpart in eq.(\ref{condo}), with  coupling constant and temperature dependent corrections,  as was also discussed in section \ref{tsebb} above.
By carrying out numerical calculations analogous to those in \cite{hanadaprobe} one can  try to determine the effective potential and equating these results  to the  potential   obtained in the bulk for a probe brane, one can further  hope to obtain the correct target space constraint corresponding to a bulk region. Our proposal would then be that the sub-algebra for this possibly modified target space constraint is the correct one to use for obtaining the bulk entanglement. Hopefully, further developments, especially in numerical methods will  lead to  concrete checks for our ideas and will allow them to be sharpened further. In fact, connecting with some of these recent developments has been one of our major motivations. 

Our proposal for associating a target space constraint with a bulk region is most straightforward when the bulk region is bounded by a surface with a constant value of one of the cartesian coordinates. This is because the fields in the Dp brane field theory as written in (\ref{three-one}) directly relate to cartesian coordinates in the geometry. For many physical questions, however, one would like to consider subregions which are bounded by constant values of the radial coordinate. A natural guess for the corresponding target space constraint is to require that the eigenvalues of the hermitian operator $\sum (X^I)^2$ are restricted to larger than some value. With this in mind, we have discussed in section \ref{gaugeinv2.3} in some detail how to implement a constraint in the radial direction in the bulk.  We  have also derived path integral expressions for Renyi entropies in a gauge fixed description which can be directly used in numerical calculations.

There are several open questions which merit further study.
As we have discussed above we expect  that  the sub -algebra of operators  we are considering will allow one  to determine the one- point function of the metric and some other supergravity fields in the vacuum and in excited coherent states.  But we do not expect to be able to determine   all  correlators  of supergravity modes in the bulk region of interest from the sub-algebra, in general. How much information can be extracted  from the sub-algebra and how does this  contrast with  the  measurements which bulk observers can do using supergravity probes restricted to the region of interest, is an important issue which needs to be understood better.
One would hope that bulk regions whose boundary is given by an RT extremal surface should correspond to rather special constraints in target space \footnote{We are thankful to Shiraz Minwalla for emphasising this point.}. Unfortunately we do not see any evidence for this so far and leave it as an important question for further investigation.  On a related note, one would think that area of extremal surfaces not of the RT type, as in D0 brane geometry \cite{vanrams} would also have some understanding in terms of target space entanglement entropy.
Finally,  for usual $AdS/CFT$ duality there is evidence in favour of the conjecture that there is an intimate connection between  entanglement in base space and emergence of a smooth AdS bulk with locality \cite{rams2} . One would expect that in models of $AdS \times {\rm (Sphere)} / CFT$ there should be a similar connection between entanglement in color space and locality in the sphere factor of the bulk. Target space entanglement provides a concrete framework to study this connection. In particular in D0 brane holography there is no base space of the holographic theory: target space entanglement would entirely account for bulk locality. This deep connection is well worth understanding further as well.





\section{Acknowledgements} We thank Masanori Hanada, Antal Jevicki, Shiraz Minwalla, Mark van Raamsdonk  and especially Suvrat Raju, for discussions. S.R.D would like to thank Tata Institute of Fundamental Research for hospitality during numerous extended visits over the years. The works of S.R.D. and S.L. have been partially supported by National Science Foundation grant NSF/PHY-1818878. The work of S.L. is supported in part by a Macadam Fellowship.
A.K., G. M. and S. P. T. acknowledge   the support of the Govt. Of India, Department of Atomic Energy, under Project No. 12-R\&D-TFR-5.02-0200 and support from the Quantum Space-Time Endowment of the Infosys Science Foundation. S. P. T. acknowledges support from a J. C. Bose Fellowship, Department of Science and Technology, Govt. of India.

\appendix

\section {Details of construction of sub-algebras}
\label{appone}

This appendix provides some details of the gauge invariant construction of the subalgebra described in Section \ref{gaugeinv}.

\subsection{Single Matrix}
First consider gauged matrix quantum mechanics of a single $N \times N$ matrix $M$.
Gauge invariant single trace operators are of the form 
\ben
\hC = {\rm Tr} \left( \hM^m \hpiM^n \right)_{\rm{order}}
\label{four-one}
\een
Here $\hpiM$ denotes the conjugate momentum to $\hM$ and the notation $\left(\right)_{\rm{order}}$ means that the $\hM$ and $\hpiM$'s are sprinkled in all possible orders. However we can use the commutation relations to bring e.g. all the $\hM$'s together. The trace Tr is over matrix indices. 

We want to construct a sub-algebra of operators which can be used to make measurements in a region $A$ of the space of eigenvalues of $\hM$. This is achieved by defining the projector (\ref{20-5}). Then the operator which will belong to this algebra is of the form
\ben
\hC_A = {\rm Tr} \left( \hP_A M \hP_A M \cdots \hP_A M \hP_A \hpiM \hP_A \cdots \hpiM \hP_A \right)
\label{four-three}
\een
In writing (\ref{four-three}) we have used $\hP_A^2 = \hP_A$ and the fact that in this particular case $\left[ \hP_A , \hM \right] = 0$.
The operators of the form (\ref{four-three}) together with the identity operator form a sub-algebra of operators of the theory. 

To see that gives us the right sub-algebra, fix a gauge where $\hM$ is diagonal with its diagonal elements denoted by $\hl_i, i= 1\cdots N$, while the diagonal elements of the conjugate momenta are denoted by $\hpi_i$. As explained in Section \ref{gaugeinv2.1}, the resulting constraint requires the states to be singlets. The remaining gauge freedom of Weyl transformations needs to be imposed by Weyl symmetrizing the states, and absorbing the standard van der Monde factor then makes $\hl_i$ coordinate operators of $N$ fermions on a line. Let us begin by considering operators which do not contain the conjugate momenta, i.e.
\ben
\cO_n = {\rm Tr} (\hM^n)
\label{foura-1}
\een
The expectation value of the operator $\cO_n$ in some pure state $|\Psi \rangle$ is given by
\ben
\sum_{i=1}^N \langle \Psi | \hl_i^n |\Psi \rangle
\label{foura-2}
\een
This measures the position of each of the fermions, takes its $n$-th power and then sums over all the particles. Similarly, the expectation value of ${\rm Tr} (\hpi_M^n )$ would measure the sum of $n$-th power of the momenta of all the fermions. This is of course a standard measurement in a system of many identical particles. 
The projected version of the operator (\ref{foura-1}) is 
\ben
\cO^P_n = \sum_{i=1}^N \int_A [ \prod_{s=1}^n dx_s] [\prod_{s=1}^n \delta (x_s - \hl_i) ]\hl_i^n
\label{four-3a}
\een
We will use the basis
\ben
|\lambda_1,\lambda_2,\cdots \lambda_N>_a = \frac{1}{N !} \sum_{\sigma \in S_N} {\rm sgn}(\sigma) | \lambda_{\sigma(1)}, \cdots  \lambda_{\sigma(1)} \rangle
\een
where $| \rangle_a$ denotes an anti-symmetrized ket.
Note that this is {\em not} an eigenstate of each individual term in the sum in (\ref{four-3a}). However it {\em is} an eigenstate of the sum. This follows from the fact that the sum is symmetric under permutations. 

Consider first the case $N=2$. The expectation value of the operator $\cO^P_3$ in a state with a wavefunction $\Psi (\lambda_1,\lambda_2)$ is given by
\begin{align}
\langle \Psi | \cO^P_3 |\Psi \rangle = \int_A [dx_1dx_2dx_3] \int_R d\lambda_1 d\lambda_2 ~\Psi^\star( & \lambda_1,\lambda_2)\{\lambda_1^3~ \delta (x_1-\lambda_1)\delta(x_2-\lambda_1)\delta(x_3-\lambda_1)\nonumber \\
& + \lambda_2^3~\delta (x_1-\lambda_2)\delta(x_2-\lambda_2)\delta(x_3-\lambda_2)\} \Psi(\lambda_1,\lambda_2)
\label{four-5a}
\end{align}
where the wavefunction is $\Psi[\lambda_1 \cdots \lambda_N] = \langle\{ \lambda_i \} | \Psi \rangle$.
where we have used antisymmetry of the wavefunctions. The integrals over $\lambda_1$ and $\lambda_2$ are over the entire real line. Writing each of these integrals as a sum over an integral over $A$ and an integral over the complement $\bA$, and noting that the delta functions ensure that only the integrals over $A$ contribute, it is straightforward to see that the result is
\begin{align}
\langle \Psi | \cO^P_3 |\Psi \rangle = & \int_A d\lambda_1\int_A d\lambda_2 \Psi^\star (\lambda_1,\lambda_2) (\lambda_1^3 + \lambda_2^3) \Psi (\lambda_1,\lambda_2) \nonumber \\
& 
+ 2 \int_A d\lambda_1\int_{\bA} d\lambda_2 \Psi^\star (\lambda_1,\lambda_2) (\lambda_1^3) \Psi (\lambda_1,\lambda_2)
\label{four-6a}
\end{align}
The first term is the contribution from configurations when both the particles are in the region of interest, while the second term from configurations where one of the particles is in the region of interest. Clearly this expectation value is equal to the expectation value of the operator without the projection if the wavefunction is non-vanishing only when {\em both} the particles are in the region of interest $A$.

This result can be easily generalized for arbitrary $N$. Then
the expectation value of the operator $\cO^P_n$ becomes
\ben
\langle \Psi | \cO^P_n |\Psi \rangle = \sum_{k =1}^{N}~{N \choose k} \sum_{a=1}^k \int_A \prod_{a=1}^k d\lambda_a \int_{\bA} \prod_{\alpha=k+1}^N d\lambda_\alpha ~\left( \Psi^\star [\{\lambda_i \}] ~ \lambda_a^n ~\Psi [\{\lambda_i \}] \right)
\label{four-4}
\een
The expression (\ref{four-4}) is a sum over sectors specified by the number of the $\lambda_i$'s in the region of interest. The expectation value then measures the sum of the $n$-th power of the position of all particles {\em which are in the region of interest}. Equation (\ref{four-4}) is simply a reflection of the decomposition of the Hilbert space into sectors, as in (\ref{four}).

The action of our projected operator on a basis state is
\ben
\cO^P_n |\{\lambda_i \} \rangle_a = \sum_{i=1}^N \int [ \prod_{s=1}^n dx_s] [\prod_{s=1}^n \delta (x_s - \lambda_i) ]\lambda_i^n |\{\lambda_i \} \rangle_a
\label{four-4a}
\een
Suppose the state is in the $(k,N-k)$ sector, i.e. $k$ of the $\lambda_i$'s lie in the region of interest. We can choose these to be the $\lambda_a, a = 1 \cdots k$. Consider a term in the sum in  (\ref{four-4a}). This contains a product of delta functions, so this will be nonzero only when the corresponding $\lambda_i$ lie in the region of interest $A$. This means that the sum over $i$ is truncated to the first $k$ terms,
\ben
\cO^P_n |\{\lambda_i \} \rangle_a = \sum_{i=1}^k \lambda_i^m |\{\lambda_i \} \rangle_a
\label{four-7a}
\een
Thus this operator acting on a basis state in the $(k,N-k)$ sector has a trivial action on the eigenvalues which are in the complement $\bA$. This is an example of an operator of the type $(k,N-k)$, 
\ben
{\cal O}_{k,N-k}~ |\{ \lambda_a\},\{\lambda_\alpha \} \rangle_a = \int \prod_{a=1}^k [d\lambda_a^\prime] ~ \tO \left( \{  \lambda_a^\prime \} ,  \{\lambda_a \} \right) ~|\{ \lambda_a^\prime \}, \{ \lambda_\alpha \} \rangle_a
\label{five}
\een
Let us define a smaller Hilbert space of $k$ particles which s spanned by
\ben
| \{ \lambda_a \}\rangle_a~~~~~~~~a=1 \cdots k ~~\lambda_a \in A
\label{four-4b}
\een
Then one can define an operator in this smaller Hilbert space,
\ben
{\tilde{\cO}}_{k,N-k} = \int [\prod_{a=1}^k d\lambda_a d\lambda_a^\prime]~\tO_{k,N-k} \left( \{ \lambda_a\},\{ \lambda_a^\prime \} \right)~
|\{\lambda_a \} \rangle_a  ~_a\langle \{ \lambda_a^\prime \} |
\label{six}
\een
In the above discussion the sector $(0,N)$ did not enter in the expression (\ref{four-4}). This simply reflects the fact that if we measure any operator involving  the position and momenta in the region of interest $A$, we should get a non-zero answer only if there are particles in $A$. However since the identity operator is also a member of the sub-algebra, this sector needs to be included. In fact the identity operator is the only operator which will receive contributions from the $(0,N)$ sector.

Clearly the expression (\ref{four-4}) can be written as a sum over traces in the smaller Hilbert spaces,
\ben
\langle \Psi | \cO^P |\Psi \rangle ={ N \choose k} \sum_{k=1}^{N-1} {\rm Tr}_A \left[ \trho_{k,N-k} \cO_{k,N-k} \right]
\label{four-5}
\een
where the density matrix $\trho_{k,N-k}$ is given by
\ben
\trho_{k,N-k} = {N \choose k} \int_A [\prod_{i=1}^k d\lambda_a d\lambda_a^\prime]\int_{\bA} [\prod_{\alpha=k+1}^N  d\lambda_\alpha]~\Psi^\star[ \{ \lambda_a \}, \{\lambda_\alpha \} ]~\Psi^\star[ \{ \lambda_a^\prime \}, \{\lambda_\alpha \} ]
|\{\lambda_a \} \rangle_a  ~_a\langle \{ \lambda_a^\prime \} |
\label{eight-1}
\een
Note that this is an operator which lives in the $k$-particle sector of the small Hilbert space defined in (\ref{four-4b}). 
This decomposition of the whole hilbert space into sectors is exactly what appears in \cite{Das:2020jhy}. The projected operators therefore provide a gauge invariant formulation of the problem.

The above constructions easily generalize to the situation when the state of the entire system is a mixed state.
Let the density matrix of the whole system be
\ben
\rho_{tot} = \int [\prod_{i=1}^N d\lambda_i d\lambda_i^\prime]~\rho_{tot} \left( \{ \lambda_i\},\{ \lambda_i^\prime \} \right) ~
|\{\lambda_i \} \rangle_a  ~_a\langle \{ \lambda_i^\prime \} |
\label{seven}
\een
Then the reduced density matrix $\trho$ which evaluates expectation values of operators belonging to this subalgebra is obtained by tracing over $\bA$,
\ben
\trho_{k,N-k} = {N \choose k}\int [\prod_{i=1}^k d\lambda_a d\lambda_a^\prime][\prod_{\alpha=k+1}^N d\lambda_\alpha]~\rho_{tot} \left( \{ \lambda_a \}, \{ \lambda_\alpha \};\{ \lambda_a^\prime \}, \{\lambda_\alpha \} \right)~
|\{\lambda_a \} \rangle_a  ~_a\langle \{ \lambda_a^\prime \} |
\label{eight}
\een
This density matrix is not normalized. In fact the trace ${\rm tr}_k\trho_k$ is the probability of $k$ particles to be in the region of interest. Thus the density matrix in the Hilbert space which is a direct sum of all the sector is properly normalized.

The von Neumann entropy associated with the reduced density matrix ${\rm tr}\trho_k$ is given by
\ben
S_{k,A} = - {\rm tr}_{\mH_{k,N-k}}(\trho_{k,N-k} \log \trho_{k,N-k} )
\label{nine}
\een
This quantifies the entanglement between the target space region $A$ and its complement $\bA$ in this sector. Following the above steps, we can easily see that a reduced density matrix $\rho_{_{RDM}}$ based on the  gauge invariant subalgebra ${\cal A}$, defined by
  \[
    {\rm Tr}(\rho_{_{RDM}} \cO) := {\rm Tr}(\rho_{tot} \cO) \quad \forall
    \cO \in {\cal A}
    \]
    satisfies
    \[
    \rho_{_{RDM}} = \oplus_{k=0}^N \trho_{k, N-k}
    \]
The total target space entanglement entropy is then a sum over all sectors
\ben
S_A = \sum_{k=0}^N S_{k,A}
\label{ten}
\een
As shown in \cite{Das:2020jhy} this quantity satisfies the usual positivity properties and strong subadditivity. Similar sector-wise entanglement also appears in discussions of entanglement entropy in gauge theories \cite{ST,V}.

We have used a first quantized description of the system. However there is an equivalent second quantized description. In the latter, we have a conventional nonrelativistic field theory of a fermion field $\psi (\lambda,t)$ : the space of this theory is the space of eigenvalues. The target space entanglement we discussed above now becomes a conventional geometric entanglement in this field theory.


\subsubsection{Momentum operators}

Operators involving momenta are subtle and at a first sight, appear to require introduction of other sectors. This can
be illustrated by a calculation of the expectation value of the projected version of an operator of the form (\ref{four-one}) with $m=0$ and $n=2$, with the region of interest $A$ being the positive real line $\lambda \geq 0$. The result of a calculation analogous to (\ref{four-6a}) is, for $N=2$,
\begin{equation}
\begin{split}
& \langle \Psi  | \operatorname{tr} ( \hat{P} \hat{\Pi}_M \hat{P} \hat{\Pi}_M \hat{P} ) | \Psi \rangle \\
=& -\frac{1}{2} \int_0^{\infty} d\lambda_1 \int_0^{\infty} d\lambda_2 \left[ \Psi^*(\lambda_1,\lambda_2) \left( \frac{\partial^2}{\partial \lambda_1^2}+\frac{\partial^2}{\partial \lambda_2^2} \right) \Psi(\lambda_1,\lambda_2) +\Psi(\lambda_1,\lambda_2) \left( \frac{\partial^2}{\partial \lambda_1^2}+\frac{\partial^2}{\partial \lambda_2^2} \right) \Psi^*(\lambda_1,\lambda_2)   \right] \\
& - \int_0^{\infty} d\lambda_1 \int_{-\infty}^0 d\lambda_2 \left[ \Psi^*(\lambda_1,\lambda_2)  \frac{\partial^2}{\partial \lambda_1^2} \Psi(\lambda_1,\lambda_2) +\Psi(\lambda_1,\lambda_2)  \frac{\partial^2}{\partial \lambda_1^2} \Psi^*(\lambda_1,\lambda_2)   \right] \\
& + \int_{-\infty}^{\infty} d\lambda \left( \frac{\partial}{\partial \lambda_1} | \Psi (\lambda_1,\lambda) |^2 \right) \bigg|_{\lambda_1=0} 
 + 2\delta(0) \int_{-\infty}^{\infty} d\lambda  | \Psi (0,\lambda) |^2
\end{split}
\label{boundary-terms}
\end{equation}
The first and second lines of the RHS are analogous to what we got in \eq{four-6a}; the first line is the contribution from the $(2,0)$ sector, and the second line is the contribution from the $(1,1)$ sector. However, we appear to also have an extra line, the third line, which represents a sector that has one of the particles exactly at $\lambda = 0$.

Note that since these extra terms pertain to particles at the boundary, it is tied to the question of how one defines the region $A$ precisely, e.g. as an open or a closed set, or in terms of a target space lattice etc. We suggest below an alternative treatment in terms of translation operators, rather than momenta, which provide a proof of principle how these problems can be avoided.

To explain this, let us start with the case of $N=1$, that is, just the case of a single $1 \times 1$ matrix or equivalently the case of one particle. We now have just two sectors $(1,0)$ and $(0,1)$, in the first one the particle is in region $A$ (which we will again define as $x>0$) and in the second one it is outside.

Consider the traslation operator
\be
O_a = \exp[- i a \hPi]
\label{o-a}
\ee
with the action
\be
O_a | x \ran = | x +a \ran
\label{o-a-ket}
\ee
Clearly such operators can take states in $(1,0)$ to $(0,1)$ (if $a<0$) or vice versa (if $a>0$). In the following, we will take $a>0$ to be specific; a similar analysis can be carried out with $a<0$. To obtain operators acting within the sector $(1,0)$, let us use the projection $O_a \to O_a^P$. It is useful to represent \eq{o-a} in terms of the product of a large number $n$ of exponentials (as in Feynman path integrals), with $a= n \eps$. In the limit of $\eps \to 0$, each exponential can be approximated as $\exp[-i\eps \hPi] \approx 1 + (-i \eps ) \hPi$ whose projected  version is $\hP( 1 + (-i \eps ) \hPi )\hP  \approx \hP( \exp[-i \eps\hPi] )\hP$. In short, we have  
\begin{align}
  O_a & \equiv \left(\exp[-i\eps \hPi]\right)^n
  \nonumber\\
  O_a^P &=  \lim_{\eps\to 0} \left(\hP \exp[-i\eps \hPi] \hP \right)^n
\nonumber
\end{align}
With the above expressions, it is easy to see that
\begin{align}
  \lan x' | O_a | x \ran  &=  \lan x'  | x+a \ran = \delta(x'- a-x)
  \label{plain-delta}\\
  \lan x' | O_a^P | x \ran  &=\lim_{\eps\to 0}  \lan x' | \left(\hP \exp[-i\eps \hPi] \hP \right)^n | x \ran
  \nonumber\\
  & = \lim_{\eps\to 0}  \theta(x') \lan x'-\eps | \theta(x'-\eps)  \left(\hP \exp[-i\eps \hPi] \hP \right)^{n-1} | x \ran
   \nonumber\\
   & = \lim_{\eps\to 0}  \theta(x')  \theta(x'-\eps)\theta(x'- 2\eps)...\theta(x'-a)  \delta(x'- a-x)
   \label{theta-theta-delta}
\end{align}
Note that unless $x$ and $x'$ are both in $A$, the above matrix element vanishes. This is because, say $x$ is not in $A$ while $x'$ is in $A$, then at least $\theta(x)= \theta(x'-a)$ will vanish, making  the entire product \eq{theta-theta-delta} vanish.\footnote{We avoid here the possibility $x=0$ by assuming that the partition demarcating the region $A$ does not fall on $x=0$. This is equivalent to assuming a lattice structure of the real line such that none of the sites falls exactly on 0, which is possible for any lattice separation, however small. Presumably a similar reasoning can get rid of the boundary terms in \eq{boundary-terms} as well.} On the other hand, if both $x$ and $x'$ {\it are} in $A$, then all the theta-functions in \eq{theta-theta-delta} evaluate to 1 (since the arguments of the theta-functions are all located on a straight `Feynman' path joining $x$ and $x'$ which are both in $A$ which is convex). This leads to the original matrix element of $O$ \eq{plain-delta} which described the case with no restrictions. There are no extra terms corresponding to particles located at $x=0$.

Now, an observable corresponds to the hermitian operator is $ {\bf O}_a =O_a + c.c.$, the projected operator being ${\bf O}^P_a =O_a^P + c.c.$. Their expectation values are given as follows. For a general wavefunction $\Psi(x)$
\begin{align}
  \lan \Psi |{\bf O}_a | \Psi\ran &= \int_{-\infty}^\infty dx \left[\Psi^*(x) \Psi(x+a) + \Psi^*(x+a)\Psi(x) \right]
  \nonumber\\
 \lan \Psi |{\bf O}_a^P | \Psi\ran &= \int_{0}^\infty dx \left[\Psi^*(x) \Psi(x+a) + \Psi^*(x+a)\Psi(x) \right]
\label{n=1-works}
\end{align}
Note that the projected operator merely restricts the range of the integral to $x>0$ as it should, and does not introduce any unwarranted boundary terms, unlike in \eq{boundary-terms}, corresponding to particles located at $x=0$.

The generalization to $N>1$ can be done as follows. Consider the translation operator $O_a= \exp[-i a \tr \hPi] = \exp[-i a \sum_{m=0}^N \hPi_m ]$. This operator is obviously gauge-invariant, since it involves $\tr \hPi$. In this case the above argument for $N=1$ can be straightforwardly generalized. The position space matrix elements of $O_a^P$ again involve a string of theta functions all located along a straight line from $X=(x_1, x_2, ...)$ to $X'=(x_1 + a, x_2 + a, ...)$, which all evaluate to 1 if both $X$ and $X'$ are in region $A$, i.e. both $x_1$ and $x_1 +a$ are positive. 

Now the reader may justifiably point out that this is not the most general translation operator, since the above operator translates the point $(x_1, x_2, ...)$ by the same amount. This can be remedied by considering an operator $\exp[-i \tr (A \hPi)]$ which evaluates to $\exp[-i \sum_{m=0}^N a_{mm} \hPi_m ]$. This clearly describes a most general translation. The operator is not gauge invariant, however, since $A$ is a fixed matrix. To make it gauge invariant, one can sum over terms with Weyl-copies of $A$.\footnote{In the previous paragraph, $A= a {\bf 1}$ which was automatically Weyl-invariant.}

We made these arguments in the context of a single matrix, but it is generalizable to multiple matrices too.

The above considerations provide a proof of principle that if we replace the momentum operators by appropriately defined `translation' operators, then the problem pointed out at the beginning of this subsubsection can be taken care of. 

\subsection{Multiple Matrices}

For multiple matrices we have two possible subalgebras which correspond to a given target space constraint.

A projector leading to the first sub-algebra is defined by (\ref{proms}), and the procedure to construct operators which belong to the sub-algebra is explained in section \ref{gaugeinv2.2}.The gauge choice which makes the physics most transparent is the one where the hermitian matrix $f(\hX^I)$ is chosen to be diagonal.
In this subsection we will discuss the simplest constraint where the function which appears in (\ref{proms}) is
\ben
F[\hX] = X^1
\label{four-two-2}
\een
As in the single matrix example, we fix a $A_t=0$ gauge and fix the remaining time independent gauge freedom by choosing $\hX^1$ to be diagonal with diagonal elements are $\hl_i$. The remaining symmetries are Weyl transformations which permute the eigenvalues and the matrix elements of the other matrices, and $U(1)^N$ transformations as in (\ref{permute-x}) and (\ref{u1}).
This symmetry is imposed by hand by adding the transforms in the states, as in (\ref{four-eleven}). As in the single matrix case, this is an eigenstate of the traced operators of the form (\ref{opproj}). Thus when ${\hat{\cO}}^{P_1}$ acts on such a state, we can replace the operators appearing in (\ref{opproj}) by their eigenvalues which we denote by the matrix without a hat. 

Acting on a state where the $\lambda_i$ for $i = 1 \cdots k$ are in the region A, the projected version of the operator $\hX^I$ as defined in (\ref{changever1})
\[
 (X^I)^{P_1}_{ij} = \int dx_1 \delta (x_1 - \lambda_i) X^I_{ij} \int dx_2 (x_2 - \lambda_j) =
  \begin{cases} 
   X^I_{ij} & \text{if } i,j = 1 \cdots k\\
   0      & \text{if } ~~{\rm otherwise}
  \end{cases}
\]
Thus the projector projects each of the matrices to the $k \times k$ block, as depicted in (\ref{matrixpi}).

Consider, for example, an operator ${\hat{\cO}}$ (as in (\ref{ctrace}) which is of the form $\cO = {\rm Tr} (\hX^I \hX^J \hX^K)$. The action of its projected version on a basis state is given by
\begin{align}{\hat{\cO}}^{P_1}
& ~| \{ \lambda_a \}, \{ \lambda_\alpha \}\{ X^L_{ab} \} \{ X^L_{a\alpha} \} \{ X^L_{\alpha a} \} \{ X^L_{\alpha\beta} \} \rangle_W  \nonumber \\
& =  \sum_{a_1\cdots a_3=1}^k  X^I_{a_1a_2} X^J_{a_2 a_3} X^K_{a_3 a_1}~
| \{ \lambda_a \}, \{ \lambda_\alpha \}\{ X^L_{ab} \} \{ X^L_{a\alpha} \} \{ X^L_{\alpha a} \} \{ X^L_{\alpha\beta} \} \rangle_W 
\label{four-twelve}
\end{align}
This is an example of an operator of type $(k,N-k)$ in the first proposal for a sub-algebra in \cite{Das:2020jhy}. An operator belonging to this first sub-algebra has a non-trivial action only on the $\lambda_a, X^L_{ab}, L \neq 1$ for $a,b = 1 \cdots k$, as shown in (\ref{fiveb}). The reduced density matrix which evaluates the expectation values of these operators is then obtained from the density matrix $\rho_{tot}$ of the whole system by tracing over the variables $\lambda_\alpha, X^I_{a\alpha},X^I_{\alpha\beta}$. This expression is given in (\ref{three-nine})

A second sub-algebra was also defined which retains the off diagonal matrices of the type $X^L_{a\alpha}$. This is defined in equations (\ref{orthodox}) - (\ref{ver2op})
Acting on a state of the form (\ref{four-eleven}) where $\lambda_i, i=1 \cdots k$ we then have
\[
(X^I)^{P_2}_{ij} =
\begin{cases}
X^I_{ij} & \text{if}~~~ i= 1\cdots k, j=1 \cdots N \\
X^I_{ij} & \text{if}~~~ i= 1\cdots N, j=1 \cdots k \\
0 & \text{otherwise}
\end{cases}
\]
It is now straightforward to see that a projected operator has a non-trivial action only on $\lambda_a,X^I_{ab},X^I_{a\alpha}$ and $X^I_{\alpha a}$. 
This is an operator of type $(k,N-k)$ in the second subalgebra defined in \cite{Das:2020jhy}, whose action is given in (\ref{fiveg}) The corresponding reduced density matrix is given in (\ref{three-nine-1})


\section{Polar Decomposition of Matrices}
\label{apptwo}

In this appendix we provide the details of the polar decomposition of multiple matrices.

\subsection{Two Matrices}

\label{apptwo_1}

For two matrices, the positive semi-definite matrix $\hR$ given by (\ref{9-4a}) is expressed in terms of unitary matrices $\hV,\hW$ by (\ref{9-5}). 
The inverse of the direct product matrix $\left[ \hat{V} \otimes \hat{V}^*   +  \hat{W} \otimes \hat{W}^* \right]$ in terms of a infinite series as follows
\begin{equation}
\left( \hat{V}^* \otimes \hat{V}   +  \hat{W}^* \otimes \hat{W} \right)^{-1} = \hat{V}^{T} \otimes \hat{V}^{\dagger} \left( \mathbb{I} \otimes \mathbb{I} + (\hat{W}\hat{V}^{\dagger})^*  \otimes \hat{W}\hat{V}^{\dagger} \right)^{-1} 
=\hat{V}^{T} \otimes \hat{V}^{\dagger}  \sum_{n=0}^{\infty} (-1)^n\left[ (\hat{W}\hat{V}^{\dagger})^*  \otimes \hat{W}\hat{V}^{\dagger} \right]^n
\label{b-2}
\end{equation}
which proves (\ref{9-8}) with the matrix $\hQ$ defined in (\ref{9-9}).

Thus $\left( \hat{s}^2\right)_{kl}$ can be solved by applying the relation (\ref{9-8})
\begin{equation}
\begin{split}
\left( \hat{s}^2 \right)_{ij} =& \left[ \left( \hat{V}^* \otimes \hat{V}   +  \hat{W}^* \otimes \hat{W} \right)^{-1} \right]_{ij, kl} 2 \left( \hat{R}^2 \right)_{kl} \\
=& 2  \left[   \sum_{n=0}^{\infty} (-1)^n \hat{V}^{T} \left[(\hat{W}\hat{V}^{\dagger})^* \right]^n \otimes \hat{V}^{\dagger} \left[ \hat{W}\hat{V}^{\dagger} \right]^n  \right]_{ij, kl} \left( \hat{R}^2 \right)_{kl}  \\
= & 2   \sum_{n=0}^{\infty} (-1)^n \left[ \hat{V}^{\dagger} \left[ \hat{W}\hat{V}^{\dagger} \right]^n \right]_{ik} \left( \hat{R}^2 \right)_{kl} \left[  \left[(\hat{W}\hat{V}^{\dagger})^{\dagger}  \right]^n\hat{V} \right]_{lj}   \\
\end{split}
\label{sq_crspr}
\end{equation}
which proves (\ref{9-10}).

We now prove that the right hand side of (\ref{9-10}) is positive semi-definite. Let $\{v_i \}$ be the set of eigenvectors of $\hat{Q}$. Because $\hat{Q}$ is unitary, the eigenvalues of $\hat{Q}$ take the form
\begin{equation}
\hat{Q} v_i = e^{i \phi_i} v_i 
\end{equation}
then we find the inner product
\begin{equation}
\begin{split}
\langle v_i, 2\left\{\sum_{n=0}^{\infty} (-1)^n   \left[ \hat{Q}^{\dagger} \right]^n  \hat{R}^2   \hat{Q}^n  \right\} v_i \rangle =& 2 \sum_{n=0}^{\infty} (-1)^n \langle v_i,  \left( \hat{Q}^{\dagger} \right)^n   \hat{R}^2  \hat{Q}^n v_i \rangle = 2 \sum_{n=0}^{\infty} (-1)^n \langle \hat{Q}^n v_i,     \hat{R}^2  \hat{Q}^n v_i \rangle \\
=& 2 \sum_{n=0}^{\infty} (-1)^n  \langle v_i,     \hat{R}^2  v_i \rangle = \langle v_i,     \hat{R}^2  v_i \rangle >0 
\end{split}
\label{psd_prf}
\end{equation}
since $\hat{R}^2$ is positive semi-definite according to (\ref{9-4a}).

Since $\hat{s}$ is positive semi-definite diagonal matrix, $\left(\hat{s}^2 \right)_{ij} = \hat{s}_i^2 \delta_{ij}$, we can take the square root of both sides of (\ref{9-10}) to get
\begin{equation}
\hat{s} = \sqrt{2}  \hat{V}^{\dagger} \left\{\sum_{n=0}^{\infty} (-1)^n   \left[ \hat{Q}^{\dagger} \right]^n  \hat{R}^2   \hat{Q}^n  \right\}^{1/2} \hat{V}
\tag{\ref{9-10}}
\end{equation}
Plugging (\ref{9-10}) back into (\ref{9-4}) we obtain $\hat{Z}$,
\begin{equation}
\hat{Z} = \hat{V} \hat{s} \hat{W}^{\dagger} = \sqrt{2}   \left\{\sum_{n=0}^{\infty} (-1)^n   \left[ \hat{Q}^{\dagger} \right]^n  \hat{R}^2   \hat{Q}^n  \right\}^{1/2} \hat{V}\hat{W}^{\dagger} = \sqrt{2}   \left\{\sum_{n=0}^{\infty} (-1)^n   \left[ \hat{Q}^{\dagger} \right]^n  \hat{R}^2   \hat{Q}^n  \right\}^{1/2} \hat{Q}
\label{arcd}
\end{equation}
This proves (\ref{9-12}), where the operation ${\mathfrak{L}}_{\hV}$ is defined in (\ref{9-13}).
In terms of matrix elements, the equation (\ref{9-12}) can be written as
\begin{equation}
\hat{Z}_{ij} = \sqrt{2} \left\{ \left[ \left( \mathbb{I} \otimes \mathbb{I} + \hat{Q}^{T} \otimes \hat{Q}^{\dagger} \right)^{-1} \right]_{il,mn}\left( \hat{R}^2 \right)_{mn}  \right\}^{1/2} \hat{Q}_{lj} .
\label{arcd_c}
\end{equation}
The identity (\ref{9-13a}) follows from the definition (\ref{9-13}),
\begin{equation}
\hat{V}^{\dagger} \left( \mathfrak{L}_{\hat{V}} \hat{M} \right)^2 \hat{V} + \left( \mathfrak{L}_{\hat{V}} \hat{M} \right)^2  = -2 \sum_{n=1}^{\infty} (-1)^n   \left( \hat{V}^{\dagger} \right)^n   \hat{M}^2  \hat{V}^n  +2\sum_{n=0}^{\infty} (-1)^n   \left( \hat{V}^{\dagger} \right)^n   \hat{M}^2  \hat{V}^n 
= 2\hat{M}^2,
\tag{\ref{9-13a}}
\end{equation}
we obtain (\ref{9-12}).

We now explain the derivation of the integration measure (\ref{9-17}). In the gauge where $\hR$ is diagonal, as in (\ref{9-15}), the expression for the complex matrix $\hZ$ is given by (\ref{9-10}) with $\hR^2$ replaced by $\hr^2$. In a Hilbert space basis where $\hr_i$ and $\hQ_{ij}$ are diagonal, the measure is given by (\ref{9-16}). Parametrizing $Q$ as in (\ref{9-17}) we have
\begin{equation}
\begin{split}
dQ =U\left( d e^{i\Phi} - \left[ e^{i \Phi }, U^{\dagger} dU \right] \right) U^{\dagger}
\end{split}
\end{equation}
where we have used the identity $dU^{\dagger} = - U^{\dagger} dU U^{\dagger}$. Then define
\begin{equation}
d S \equiv U^{\dagger} dU
\tag{\ref{9-18}}
\end{equation}
We can see that $d S^{\dagger}  = - dS$. 
Then the line element becomes
\begin{equation}
\begin{split}
\operatorname{tr} \left(dQ dQ^{\dagger} \right) 
=&  \operatorname{tr} \left\{ de^{i\Phi} de^{-i \Phi} + \left[ e^{i \Phi }, dS \right]\left[ e^{-i \Phi }, dS \right] \right\} \\
=& \sum_i d\phi_i^2 +2 \sum_{ij} e^{i \phi_i }dS_{ij} e^{-i\phi_j} dS_{ji} - 2 \sum_{ij} dS_{ij}dS_{ji} \\
=& \sum_i d\phi_i^2 + 8 \sum_{i < j} \sin^2 \frac{\phi_i-\phi_j}{2}dS_{ij} dS_{ij}^* \\
\end{split}
\tag{\ref{9-19}}
\end{equation}
This implies that we can choose $dS_{ii} =0$. Now we have metric $ds^2= g_{AB} d\bar{x}^A dx^B$ with $dx^A = \left( d\phi_i, dS_{ij(i<j)}, dS_{ij(i<j)}^* \right)$, where
\begin{equation}
g_{AB}= \left(
\begin{matrix}
1 & 0  & 0 \\
0 & 4 \sin^2 \frac{\phi_i-\phi_j}{2} \\
0 & 0 &  4 \sin^2 \frac{\phi_i-\phi_j}{2} \\
\end{matrix} \right)
\end{equation}
The determinant of $g$ is
\begin{equation}
\det g_{AB}= \prod_{i<j} \left( 4 \sin^2 \frac{\phi_i-\phi_j}{2} \right)^2
\end{equation}
Thus
\begin{equation}
\prod_{ij}[dQ_{ij}] = \sqrt{\det g_{AB}} \prod_i d \phi_i \prod_{i<j} dS_{ij} dS_{ij}^*
\end{equation}
The final expression (\ref{9-20}) follows when we use this in (\ref{9-16}). To ensure that the variables $r_i,\phi_i$ cover the $\mathbb{R}^2$ formed by $X^1,X^2$ once we see that the ranges of the angles $\phi_i$ are
\begin{equation}
-\pi < \phi_i < \pi, ~~~i =1, \cdots ,N
\end{equation}

\subsection{Three Matrices}

Now consider three matrices $\hat{X}^1, \hat{X}^2, \hat{X}^3$, with $\hR$ defined by (\ref{11-4}). To obtain a polar decomposition we first form a complex matrix as follows
\begin{equation}
\hat{Y} \equiv \sqrt{(\hat{X}^1)^2 + (\hat{X}^2)^2}  + i \hat{X}^3
\end{equation}
so that
\begin{equation}
2\hat{R}^2 = \hat{Y}\hat{Y}^{\dagger} + \hat{Y}^{\dagger} \hat{Y}
\end{equation}
We can now use the procedure we used for two matrices to write
\begin{equation}
\hat{Y} = \sqrt{2} \left\{   \sum_{n=0}^{\infty} (-1)^n   \left( \hat{Q}_1^{\dagger} \right)^n   \hat{R}^2  \hat{Q}_1^n   \right\}^{1/2}  \hat{Q}_1
= \left( \mathfrak{L}_{\hat{Q}_1} \hat{R} \right) \hat{Q}_1
\end{equation}
where $\hQ_1$ is a unitary matrix. 
Therefore, in manner analogous to (\ref{9-14}) we get
\begin{eqnarray}
\sqrt{(\hat{X}^1)^2 + (\hat{X}^2)^2} = \frac{\hat{Y} + \hat{Y}^{\dagger}}{2} =  \frac{ \left(\mathfrak{L}_{\hat{Q}_1} \hat{R} \right) \hat{Q}_1 + \hat{Q}_1^{\dagger} \left( \mathfrak{L}_{\hat{Q}_1} \hat{R}\right) }{2}  \label{threepsd} \\
\hat{X}^3  = \frac{\hat{Y} - \hat{Y}^{\dagger}}{2i} = \frac{ \left(\mathfrak{L}_{\hat{Q}_1} \hat{R} \right) \hat{Q}_1 - \hat{Q}_1^{\dagger} \left( \mathfrak{L}_{\hat{Q}_1} \hat{R}\right) }{2i} \label{threepsd2}
\end{eqnarray}
The next step is to consider the $\hX^1,\hX^2$ exactly as in the two matrix example in the previous subsection,
\begin{equation}
\hat{Z} \equiv \hat{X}^1 + i \hat{X}^2
\end{equation}
Then we have
\begin{equation}
(\hat{X}^1)^2 + (\hat{X}^2)^2 = \frac{\hat{Z}\hat{Z}^{\dagger}+ \hat{Z}^{\dagger}\hat{Z}}{2}
\end{equation}
Since (\ref{threepsd}) implies
\begin{equation}
(\hat{X}^1)^2 + (\hat{X}^2)^2  = \frac{1}{4} \left[ \left(\mathfrak{L}_{\hat{Q}_1} \hat{R} \right) \hat{Q}_1 + \hat{Q}_1^{\dagger} \left( \mathfrak{L}_{\hat{Q}_1} \hat{R}\right) \right]^2
\end{equation}
it is clear we need to introduce another unitary matrix $\hQ_2$ to write
\begin{equation}
\hat{Z} =\frac{1}{2}\left[ \mathfrak{L}_{\hat{Q}_2}  \left( (\mathfrak{L}_{\hat{Q}_1} \hat{R} ) \hat{Q}_1 + \hat{Q}_1^{\dagger} ( \mathfrak{L}_{\hat{Q}_1} \hat{R}) \right)\right] \hat{Q}_2
\label{13-1}
\end{equation}
This construction is exactly like (\ref{9-12}) with the replacements 
\ben
\hQ \rightarrow \hQ_2~~~~~
\hR \rightarrow \frac{1}{2}\left[  (\mathfrak{L}_{\hat{Q}_1} \hat{R} ) \hat{Q}_1 + \hat{Q}_1^{\dagger} ( \mathfrak{L}_{\hat{Q}_1} \hat{R}) \right]
\een
Finally one can express $\hX^1,\hX^2$ in terms of $\hZ$ and $Q_2$ and use (\ref{13-1}) to rewrite these in terms of $\hR, \hQ_1, \hQ_2$, while $\hX_3$ is already expressed in terms of these in (\ref{threepsd2}). This leads to (\ref{11-3}).
Finally one can check (\ref{11-4}) directly,
{
\begin{equation}
\begin{split}
(\hat{X}^1)^2 + (\hat{X}^2)^2 =& \frac{1}{2} \left[ \mathfrak{L}_{\hat{Q}_2} \frac{ \left(\mathfrak{L}_{\hat{Q}_1} \hat{R} \right) \hat{Q}_1 + \hat{Q}_1^{\dagger} \left( \mathfrak{L}_{\hat{Q}_1} \hat{R}\right) }{2}\right] ^2 + \frac{1}{2} \hat{Q}_2^{\dagger} \left[ \mathfrak{L}_{\hat{Q}_2} \frac{ \left(\mathfrak{L}_{\hat{Q}_1} \hat{R} \right) \hat{Q}_1 + \hat{Q}_1^{\dagger} \left( \mathfrak{L}_{\hat{Q}_1} \hat{R}\right) }{2}\right] ^2 \hat{Q}_2 \\
=& \left[  \frac{ \left(\mathfrak{L}_{\hat{Q}_1} \hat{R} \right) \hat{Q}_1 + \hat{Q}_1^{\dagger} \left( \mathfrak{L}_{\hat{Q}_1} \hat{R}\right) }{2}\right] ^2
\end{split}
\end{equation}}
Thus
\begin{equation}
\begin{split}
(\hat{X}^1)^2 + (\hat{X}^2)^2 + (\hat{X}^3)^2 =&  \left[  \frac{ \left(\mathfrak{L}_{\hat{Q}_1} \hat{R} \right) \hat{Q}_1 + \hat{Q}_1^{\dagger} \left( \mathfrak{L}_{\hat{Q}_1} \hat{R}\right) }{2}\right] ^2 + \left[\frac{ \left(\mathfrak{L}_{\hat{Q}_1} \hat{R} \right) \hat{Q}_1 - \hat{Q}_1^{\dagger} \left( \mathfrak{L}_{\hat{Q}_1} \hat{R}\right) }{2i} \right]^2 \\
=& \frac{1}{2} \left( \mathfrak{L}_{\hat{Q}_1} \hat{R} \right)^2 + \frac{1}{2} \hat{Q}_1^{\dagger}\left( \mathfrak{L}_{\hat{Q}_1} \hat{R} \right)^2 \hat{Q}_2 = \hat{R}^2
\end{split}
\tag{\ref{11-4}}
\end{equation}

To find the ranges of integration let us now work in a Hilbert space basis which are eigenstates of $\hat{r}_i$ and the $(\hat{Q}_1)_{ij}, (\hat{Q}_2)_{ij}$ with eigenvalues $r_i, (Q_1)_{ij}, (Q_2)_{ij}$. Unlike the case of two matrices we now have additional constraints on $Q_1$.
This is because $\sqrt{({X}^1)^2 + ({X}^2)^2} $ should be positive semi-definite. That is,
\begin{equation}
\frac{ \left(\mathfrak{L}_{{Q}_1} {R} \right) {Q}_1 + {Q}_1^{\dagger} \left( \mathfrak{L}_{{Q}_1} {R}\right) }{2} >0
\end{equation}
Now let the set of eigenvectors of $Q_1$ be $\{ v_i \}$ so that $Q_1 v_i = e^{i (\phi_1)_i} v_i$. Then
\begin{equation}
\begin{split}
0 < & \frac{1}{2}\langle v_i, \left[ \left(\mathfrak{L}_{{Q}_1} {R} \right) {Q}_1 + {Q}_1^{\dagger} \left( \mathfrak{L}_{{Q}_1} {R}\right) \right] v_i \rangle = \frac{1}{2} \left[ \langle v_i, \left(\mathfrak{L}_{{Q}_1} {R} \right) {Q}_1 v_i  \rangle + \langle \left(\mathfrak{L}_{{Q}_1} {R} \right) {Q}_1 v_i,  v_i  \rangle \right] \\
=& \frac{1}{2}\left[ e^{i(\phi_1)_i}\langle v_i, \left(\mathfrak{L}_{{Q}_1} {R} \right)  v_i  \rangle + e^{-i(\phi_1)_i}\langle \left(\mathfrak{L}_{{Q}_1} {R} \right)  v_i,  v_i  \rangle \right] = \cos (\phi_1)_i \langle v_i, \left(\mathfrak{L}_{{Q}_1} {R} \right)  v_i  \rangle
\end{split}
\end{equation}
since $\mathfrak{L}_{{Q}_1} {R}$ is Hermitian. Then given that $\mathfrak{L}_{{Q}_1} {R}$ is positive semi-definite i.e. $\langle v_i, \left(\mathfrak{L}_{{Q}_1} {R} \right)  v_i  \rangle$, we have 
\begin{equation}
\cos (\phi_1)_i >0 
\end{equation}
for $i=1, \cdots ,N$. This means we need to restrict the range of the $(\phi_1)_i$'s
\ben
-\frac{\pi}{2} \leq (\phi_1)_i \leq \frac{\pi}{2}
\een
On the other hand, there is no condition on the eigenvalues of $Q_2$. These conditions lead to (\ref{11-6}), and the measure of integration is (\ref{11-7}).

\subsection{More Matrices}

Repeating using the strategy shown in (\ref{9-12}), we can transfer matrices $\{ \hat{X}^I \}_{I=1,\cdots,D}$ into $\{\hat{R};\hat{Q}_A \}_{A=1,\cdots,D-1}$. The transformation is similar to $D$-spherical coordinates
\begin{equation}
\begin{aligned}x_{D}&=r\sin(\varphi _{1})\\x_{D-1}&=r\cos(\varphi _{1})\sin(\varphi _{2})\\x_{D-2}&=r\cos(\varphi _{1})\cos(\varphi _{2})\sin(\varphi _{3})\\&\vdots \\x_{2}&=r\cos(\varphi _{1})\cdots \cos(\varphi _{D-2})\sin(\varphi _{D-1})\\x_{1}&=r\cos(\varphi _{1})\cdots \cos(\varphi _{D-2})\cos(\varphi _{D-1}).\end{aligned}
\label{cls}
\end{equation}
with
\begin{equation}
r\sin \varphi \to   \frac{ \left(\mathfrak{L}_{\hat{Q}} \hat{R} \right) \hat{Q} - \hat{Q}^{\dagger} \left( \mathfrak{L}_{\hat{Q}} \hat{R}\right) }{2i} ,~~~r\cos \varphi \to   \frac{ \left(\mathfrak{L}_{\hat{Q}} \hat{R} \right) \hat{Q} + \hat{Q}^{\dagger} \left( \mathfrak{L}_{\hat{Q}} \hat{R}\right) }{2}
\label{qls}
\end{equation}

In a Hilbert space basis which are eigenstates of $\hat{r}_i$ and the $(\hat{Q}_A)_{ij}$ with eigenvalues $r_i, (Q_A)_{ij}$, we still need to find out the constraints on $\{{R};{Q}_A \}_{A=1,\cdots,D-1}$. Firstly, according to the argument in section \ref{apptwo_1}, we can always choose ${R}$ to be positive semi-definite since it appears in the form of ${R}^2$. Moreover, according to (\ref{psd_prf}), we can always choose
\begin{equation}
\mathfrak{L}_{{Q}_1} {R} = \sqrt{2} \left\{   \sum_{n=0}^{\infty} (-1)^n   \left( {Q}^{\dagger} \right)^n   {R}^2  {Q}^n   \right\}^{1/2} >0
\end{equation}

Now we consider $\{{Q}_A \}_{A=1,\cdots,D-1}$. Notice that in (\ref{cls}) we have
\begin{equation}
\varphi_A \in \left\{
\begin{array}{cc}
(-\pi/2, \pi/2) & A=1,...,D-2 \\
(-\pi, \pi) & A=D-1 \\
\end{array}   
\right.
\end{equation}
to avoid counting the space repeatedly. Then in matrix case, we should have similar conclusion that
if we define
\begin{eqnarray}
{R}_{A+1} &\equiv& \frac{ \left(\mathfrak{L}_{{Q}_A} {R}_A \right) {Q}_A + {Q}_A^{\dagger} \left( \mathfrak{L}_{{Q}_A} {R}_A\right) }{2},~~~ A=1,...,D-2, \\
{R}_1 &\equiv& {R},
\end{eqnarray}
then ${R}_A, A=1,...,D-1$ are all positive semi-definite.

We use Mathematical induction to derive the constraints on ${Q}_A, A=1,...,D-2$:
\begin{enumerate}
\item ${R}_1 = {R}$ is positive semi-definite; \label{step1}
\item Assume ${R}_A$ is positive semi-definite. Then we have $\mathfrak{L}_{{Q}_A} {R}_A$ positive semi-definite according to (\ref{psd_prf}).

Now for ${R}_{A+1}$, let $\{ ({v}_A)_i \}$ be the set of eigenvectors of ${Q}_A$, i.e. ${Q}_A ({v}_A)_i ={q}_A ({v}_A)_i$. Then for the complete set formed by $\{{u}_A \}$:
\begin{equation}
\begin{split}
0 < & \langle ({v}_A)_i, {R}_{A+1} ({v}_A)_i \rangle = \frac{1}{2} \langle ({v}_A)_i, \left(\mathfrak{L}_{{Q}_A} {R}_A \right) {Q}_A + {Q}_A^{\dagger} \left( \mathfrak{L}_{{Q}_A} {R}_A\right) ({v}_A)_i \rangle  \\
=& \frac{1}{2} \langle \left(\mathfrak{L}_{{Q}_A} {R}_A \right) {Q}_A ({v}_A)_i, ({v}_A)_i \rangle +\frac{1}{2} \langle ({v}_A)_i , \left(\mathfrak{L}_{{Q}_A} {R}_A \right) {Q}_A ({v}_A)_i \rangle = \operatorname{Re} {q}_A \langle ({v}_A)_i , \left(\mathfrak{L}_{{Q}_A} {R}_A \right)  ({v}_A)_i \rangle
\end{split}
\label{ps_prf}
\end{equation}
Thus given that $\mathfrak{L}_{{Q}_A} {R}_A$ is positive semi-definite i.e. $\langle ({v}_A)_i , \left(\mathfrak{L}_{{Q}_A} {R}_A \right) ({v}_A)_i \rangle >0$, we have $\operatorname{Re} {q}_A >0 $ i.e. ${Q}_A$ is positively stable. Because the eigen-basis of ${Q}_A$ forms a complete set, when ${R}_{A+1}, {R}_A, {Q}_A$ are all $N \times N$ matrices, it should be a necessary and sufficient condition. \label{step2}
\end{enumerate}

From \ref{step1}$^{\circ}$ and \ref{step2}$^{\circ}$, we can show that ${Q}_A, A=1,...,D-2$ should be positively stable. Because ${Q}_A$ are unitary matrices, their eigenvalues have the form
\begin{equation}
Q_A \equiv U_A e^{i \Phi_A} U_A^{\dagger}, ~~~\Phi_A = \operatorname{diag}[ (\phi_A)_1, (\phi_A)_2..., (\phi_A)_N]
\end{equation}
Thus "positively stable" means that 
\begin{equation}
\operatorname{Re} e^{i (\phi_A)_i}=\cos (\phi_A)_i >0,~~~ i=1,...,N; A=1,...,D-2
\end{equation}
The conditions of $\Phi^A, A=1,...,D-1$ are 
\begin{equation}
\phi^A_i \in \left\{
\begin{array}{cc}
(-\pi/2, \pi/2) & A=1,...,D-2 \\
(-\pi, \pi) & A=D-1 \\
\end{array}   
\right.~~~i=1,...,N
\tag{\ref{angle}}
\end{equation}

As a result, the measure of integration is
\begin{equation}
\begin{split}
\left[\prod_{I}^D d X^I \right] =  \mathbb{J} (r_i, (\phi_A)_i, (S_A)_{ij}) \prod_i dr_i  \prod_{A=1}^{D-1} \left[ \prod_i d(\phi_A)_i  
\prod_{i < j} [ 4   \sin^2 (\frac{(\phi_A)_i - (\phi_A)_j}{2}) d(S_A)_{ij} d(S_A^\star)_{ij} ] \right]
\end{split}
\label{jb_gr}
\end{equation}
where
\ben
dS_A \equiv U_A^\dagger dU_A, ~~~A=1,...,D-1
\een


\section{DBI+CS action for probe D0 brane}
\label{appthree}

Consider a probe D0 brane moving in the near-horizon background (\ref{ten-2}) produced by a stack of $N$ other D0 branes. The action is given by the Dirac-Born-Infeld and Chern-Simons action. In the static gauge this is given by
\ben
S = -\frac{1}{g_s l_s} \int dt \left[ e^{-\phi} \sqrt{-g_{00}-g_{IJ}\dot{x}^I \dot{x}^J} + 2A_0 \right]
\label{ac-1}
\een
where the metric $g_{\mu\nu}$, the dilaton $\phi$ and the 1-form gauge fields are given in (\ref{ten-2}). Defining the velocity $v$ by
\begin{equation}
v^2 \equiv \delta_{IJ} \partial_t x^I \partial_t x^J
\end{equation}
we expand the action in powers of $v$. This gives
\ben
\begin{split}
S 
= & \int dt \left[ \frac{1}{2R_s} v^2 +\frac{15}{16}\frac{N}{R_s^3 M_p^9}  \frac{v^4}{r^7}  +\frac{225}{64}\frac{N^2}{R_s^5 M_p^{18}}   \frac{v^6}{ r^{14}} + \cdots  \right] -\frac{1}{R_s}\int dt
\label{ac-2}
\end{split}
\een
where we have used (\ref{ten-1}) and (\ref{eleven-1}) and expressed the coefficients in terms of M theory quantities
\ben
R_s = g_s l_s~~~~~~\ell_p=g_s^{1/3}l_s~~~~~~M_p^{-9}=  ( 2\pi )^3 \ell_p^9 = (2\pi)^3 g_s^3 l_s^9
\label{ac-3}
\een
The action (\ref{ac-2}) is in precise agreement with the action of a 11 dimensional graviton with light cone momentum $p_-=1/R_s$ in the presence of another graviton with momentum $p_- = N/R_s$. The same action is obtained from the matrix theory calculation. For more details of the latter calculation see \cite{bbs}, section 12.2.

\end{document}